\documentclass[
 preprint, 
nofootinbib,
 amsmath,amssymb,
 aps, physrev, showkeys,
]{revtex4-2}

\usepackage{graphicx}
\usepackage{dcolumn}
\usepackage{bm}
\usepackage{xcolor}
\usepackage{lipsum}
\usepackage{subcaption}
\usepackage{float}
\usepackage{afterpage}
\usepackage{hyperref}
\usepackage{upgreek}
\usepackage{amsmath}
\DeclareMathOperator{\sign}{sign}

\DeclareCaptionSubType[alph]{figure}
\makeatletter
\renewcommand\p@subfigure{\thefigure~}
\makeatother

\definecolor{ifblue}{RGB}{0, 51, 102}
\definecolor{ifred}{HTML}{61172b}
\definecolor{ifgreen}{RGB}{0, 180, 0}

\newcommand{\p}[1]{
\left({#1}\right)}
\newcommand{\I}{{\cal I}}

\begin{document}


\title{
    \textbf{  Topological charges and confined-deconfined phase 
    transition in holography}}

\author{Nelson R. F. Braga}
\email{braga@if.ufrj.br}
\affiliation{Instituto de Física, Universidade Federal do Rio de Janeiro, Caixa Postal 68528, RJ 21941-972, Brazil.}

\author{William S. Cunha}
\email{wscunha@pos.if.ufrj.br}
\affiliation{Instituto de Física, Universidade Federal do Rio de Janeiro, Caixa Postal 68528, RJ 21941-972, Brazil.}


\date{\today}

\begin{abstract}
In recent years, many interesting works providing a topological description for black hole (BH) properties have appeared in the literature.  In particular, in this framework BHs correspond to topological defects in an enlarged (off-shell) parameter space, with an associated total topological charge. In gauge/gravity duality the transition from the confined to the deconfined phase is mapped into the dominance of a BH phase in the gravity side. Here we show, using a holographic AdS/QCD model, that the introduction of an energy scale in anti-de Sitter (AdS) space results in a change in the topological class. Such a modification corresponds to the existence of confined and deconfined phases, separated by a Hawking Page transition at a finite critical temperature. 
\end{abstract}

\maketitle
\newpage

\section{\label{sec: Introdução} Introduction}

     Holographic models inspired by the AdS/CFT correspondence are a very useful framework for studying strong interactions in non-perturbative regimes. These models break conformal invariance through the introduction of energy scales related to hadronic phenomenology. They provide not only spectra of masses and decay constants in the vacuum, but also describe the hadronic behavior in a thermal medium and other properties. Some examples are found in Refs.~\cite{Dudal:2015wfn,Braga:2017oqw,Braga:2019yeh,Braga:2019xwl,Braga:2021fey,Ballon-Bayona:2024yuz,Nascimento:2025smf}.

    As discussed in Ref.~\cite{Witten:1998zw}, at finite temperature the gravitational sector of the AdS/CFT duality contains two competing geometries. One is an AdS black hole space that corresponds to a deconfined phase of the dual gauge theory and dominates for temperatures above some critical value. The other is a thermal AdS space, which dominates for lower temperatures and represents the confined phase. The Hawking-Page (HP) transition \cite{Hawking:1982dh} between these geometries represents a transition of the phases of the boundary gauge theory. 
    In the AdS/CFT case, the transition occurs at a non-vanishing temperature only when the boundary of the space is compact\cite{Witten:1998zw}. It is possible to describe a plasma with a finite critical temperature in a non-compact space by incorporating an energy scale into the gravitational action. This way, one builds holographic models that make it possible to study the thermodynamics of the transition between the confined (hadronic) phase and the quark-gluon plasma (QGP) \cite{Herzog:2006ra, BallonBayona:2007vp}. This transition has been widely explored across various scenarios, including finite chemical potential, the presence of magnetic fields, and rotation. For recent examples, see Refs.~\cite{ Braga:2022yfe,Braga:2024nnj,Braga:2025wlx}

    On the other hand, by employing Duan's \(\phi\)-mapping theory \cite{Duan:1979ucg, Duan:1984ws}, it is possible to study the black hole properties from a topological perspective (see, for instance,  \cite{Wei:2020rbh,Wei:2021vdx, Wei:2022dzw, Wei:2024gfz, Wu:2024asq, Wu:2024rmv, Wu:2025xxo}). With this purpose, a conserved topological current can be introduced as
    \begin{equation}\label{eq: topological current definition}
        j^\mu =\frac{1}{2\pi}\epsilon^{\mu\nu\rho}\epsilon_{a b} \partial_\nu \text{n}^a \partial_\rho \text{n}^b
    \end{equation}
    where \(\text{n}^a\) is the unit vector associated with a two-component time-independent field \(\phi^a\):  $ \text{n}^a = \phi^a / \vert {\vec \phi} \vert  $ . Note that the Latin indices run over the two field components (\(a, b=1,2\)), while Greek indices run over auxiliary space-time components (\(\mu,\nu,\rho=0,1,2\)). The fields do not depend on the time ($\mu = 0$) coordinate.

    By using the Jacobian tensor  \(\epsilon^{ab}J^\mu (x, \phi)=\epsilon^{\mu\nu\rho}\partial_\nu\phi^a\partial_\rho\phi^b\) and the two-dimensional Laplacian Green's function 
    \begin{equation}
        \partial_a\partial_a \ln \phi = 2\pi\delta^2(\vec \phi)\,, \quad \text{with} \quad \partial_a = \frac{\partial}{\partial \phi^a}\,,
    \end{equation}
    where \(\phi\equiv|\vec\phi|\), the topological current \eqref{eq: topological current definition} can be rewritten as 
    \begin{equation}
        j^\mu = \delta^2(\vec \phi) J^\mu(x, \phi).
    \end{equation}
    
    Recalling the properties of the \(\delta\)-function, the density of the topological current can be expressed as 
    \begin{equation}\label{eq: density current as winding number}
        j^0 = \sum_{i} w_i \delta^2(x-z_i) \,,
    \end{equation}
    where the \(z_i\) denote the zeros of the field \(\vec \phi\) and \(w_i=\beta_i \eta_i\) is the winding number. Here,  \(\beta_i\) is the Hopf index, which counts the number of loops around the zero \(z_i\) in the vector space, while \(\eta_i=\text{sign}\left.\left(J^0(x, \phi)\right)\right|_{z_i}\) is the Brouwer degree \cite{Duan:1984ws, Wei:2021vdx}.

    The topological charge is obtained by integrating the topological current density over the entire two-dimensional space \(\Sigma\). Hence,  using Eq. \eqref{eq: density current as winding number}, one obtains
    \begin{equation}\label{eq: total topologic charge definition}
        W =\int_\Sigma j^0 d^2x = \sum_i w_i
    \end{equation}
    such that the topological charge (or total topological number) is the sum of all winding numbers in the space.

    Moreover, using the definition \eqref{eq: topological current definition},  the total topological number \eqref{eq: total topologic charge definition} can be rewritten as
    \begin{align}
        W &= \int_\Sigma j^0 dS_0 =\frac{1}{2\pi}\int_\Sigma dS_0\epsilon^{0 p q} \epsilon_{a b}\partial_p\text{n}^a\partial_q \text{n}^b \nonumber\\
        &=\frac{1}{2\pi}\int_\Sigma \epsilon^{p q} \epsilon_{a b}\partial_p\text{n}^a\partial_q \text{n}^b d^2 x\,\nonumber\\
        &= \frac{1}{2\pi}\oint_{\partial\Sigma}\epsilon_{a b}\text{n}^a\partial_q\text{n}^b dx^q\label{eq: line integration of the normalized field}
    \end{align}
    where Stokes' theorem has been applied to transition from the second to the third line. Since  \( \text{n}^a  \)  is a unity vector, its components can be expressed as
    \begin{equation}
        \text{n}^1 = \cos\Omega \quad \text{and} \quad \text{n}^2 = \sin\Omega\,.
    \end{equation}

    By substituting this back into Eq. \eqref{eq: line integration of the normalized field}, the total topological number becomes 
    \begin{equation}\label{eq: winding number as variation angle}
        W = \frac{1}{2\pi}\oint_{\partial\Sigma}d\Omega = \frac{1}{2\pi}\Delta\Omega\,,
    \end{equation}
    meaning that the topological number counts the total variation of the field direction along the boundary, this discussion as detailed in \cite{Wu:2025xxo}.

    In some cases we are interested in computing the winding number \(w_i\) of a single defect located at \(x = z_i\). In such cases, the integration is restricted to a neighborhood of the zero \(z_i\), and  \(w_i\) measures the variation of the field direction along a contour \({\cal C}\) that encloses this defect.

    For this purpose, we parametrize the contour by an angle \(\vartheta \in (0,2\pi)\) using the following coordinate transformation \cite{Wei:2021vdx}
    \begin{equation}\label{eq: coordenate transformation}
        x^1 = \lambda_1 \cos\vartheta + A \quad\text{and} \quad x^2 = \lambda_2\sin\vartheta +B\,,
    \end{equation}
    where \(A\) and \(B\) are constants that denote the position of the defect, while \(\lambda_1\) and \(\lambda_2\) represent the radii of the contour. Note that these two are arbitrary, as the choice of the contour \(\cal C\)  does not affect the winding number as long as the defect remains enclosed and there is no other defect included. 

    Finally, by substituting the coordinates \eqref{eq: coordenate transformation} into Eq. \eqref{eq: line integration of the normalized field}, the winding number for a single defect can be computed as 
    \begin{equation}\label{eq: deflection angle integral}
        w_i =\frac{1}{2\pi}\int^{2\pi}_0\epsilon_{a b}\text{n}^a\partial_\vartheta \text{n}^b d\vartheta\,.
    \end{equation}

    Using this framework,  Ref. \cite{Wei:2021vdx} studied the small-large black hole phase transition by employing a field derived from the Hawking temperature. Following a similar procedure, it was shown in \cite{Yerra:2022coh} that the HP transition is a topological defect. Furthermore, in \cite{Wei:2022dzw} it was proposed that the black hole solutions themselves appear as topological defects in the parameter space of a field derived from the off-shell free energy. Consequently, each black hole can be assigned a topological classification according to its total topological charge \((W)\) \cite{Wei:2024gfz}. Some recent studies using this approach can be found in \cite{Wu:2024rmv, Wu:2025xxo, Babaei-Aghbolagh:2025qxm, Yang:2025uul}.

    In this work, we apply the topological framework to the soft-wall AdS/QCD model in order to gain insights into the confinement/deconfinement transition from a topological perspective. We begin in Sec.\ref{sec: Sec2} by classifying the black hole solutions (Schwarzschild and Reissner–Nordström) in \(AdS_{_D}\) with a non-compact boundary. In Sec. \ref{sec: Sec3},  we examine how the addition of the holographic model parameters affects the classification of \(AdS_5\) black holes and how the HP transition manifests within the parameter space dynamics. In Sec.\ref{sec: Sec4}, we employ the field derived from an effective temperature to study the global transitions, showing that the confinement/deconfinement appears as a defect with a positive topological charge, as expected.  Finally, in Sec.\ref{sec: Conclusão}, our conclusions are presented.

\section{\label{sec: Sec2} Topological classes of \(\text{BH-AdS}_{_D}\) space  with \(R^{^{D-2}} \times S^1 \) boundary}

    As stated in the Introduction, it was shown in \cite{Wei:2022dzw} that the black hole solutions appear as topological defects in the thermodynamic parameter space, where they can be classified according to their topological charges. 
    
    The BH-AdS spaces commonly used in holography applications possess a geometry in the form 
    \begin{equation}\label{eq: bh ads D metric}
        ds^2 = \frac{L^2}{z^2}\p{-f(z)dt^2 + f(z)^{-1}dz^2 + d\vec x^2_n}\,,
    \end{equation}
    where \(L\) is the AdS radius, \(n=D-2\) is the number of boundary spatial dimensions, and \(f(z)\) is the horizon function.
    Let us investigate the topological classification of these spaces in arbitrary dimensions.

    We define the field \(\phi\) in the parameter space \(\p{z_h, \Theta}\) as 
    \begin{equation}\label{eq: topological field defition 2}
        \phi(z_h, \Theta) = \p{\frac{\partial {\cal 
        F}}{\partial z_h}\,, \,- \frac{\cot\Theta}{\sin\Theta}}\,,
    \end{equation}
    where \(\Theta\in(0, \pi)\) is an auxiliary parameter, \(z_h\) is the horizon position and \({\cal F}\) the off-shell free energy, defined as 
    \begin{equation}\label{eq: off shell free energy definition}
        {\cal F} = \langle E\rangle - \bar T S\,,
    \end{equation}
    where \(\bar T\) is an independent temperature variable. The connection to \(z_h\) is established through the minimization of this free energy, thereby recovering the Hawking temperature.

      \subsection{AdS Schwarzschild}

        The first case of interest is the Schwarzschild geometry, where the horizon function in the metric \eqref{eq: bh ads D metric} takes the form
        \begin{equation}\label{eq: schwarzschild horizon function}
            f(z) = 1 - \frac{z^{n+1}}{z_h^{n+1}}\,.
        \end{equation}
        The thermal AdS space is recovered setting \(f(z)=1\), or equivalently, by taking the limit \(z_h\to\infty\). The Hawking temperature obtained from this horizon function is 
        \begin{equation}\label{eq: Schwarzschild hawking temperature}
            T = \frac{\p{n+1}}{4\pi}\frac{1}{z_h}.
        \end{equation}

        The Einstein-Hilbert action for this space is given by 
        \begin{equation}\label{eq: AdS action}
            {\cal I} = - \frac{1}{2 \kappa^2}\int d^{n+2}x\sqrt{|g|}\p{{\cal R}- \Lambda}
        \end{equation}
        where \(k^2 = 8\pi G_{n+2}\) with the \(\p{n+2}\)-dimensional Newton's constant. Here, \({\cal R}\) and \(\Lambda\) denote the Ricci scalar and the cosmological constant, respectively, which for AdS  spaces can be written in terms of \(n\) as  
        \begin{equation}\label{eq: Ricci and cosmological constant}
            {\cal R} = - \frac{\p{n+1}\p{n+2}}{L^2} \quad \text{and} \quad \Lambda = - \frac{n\p{n+1}}{L^2}\,.
        \end{equation}

        Inserting these relations into action \eqref{eq: AdS action} and using the metric \eqref{eq: bh ads D metric}, the action for the BH-AdS geometry takes the form
        \begin{equation}\label{eq: bh action}
            {\cal I}^{(BH)} = \frac{L^n}{\kappa^2}\beta V_n\p{\frac{1}{\varepsilon^{n+1}}-\frac{1}{z_h^{n+1}}}\,.
        \end{equation}
        where \(V_n\) denotes the \(n\)-dimensional spatial volume and \(\beta\) is the inverse of temperature arising from the imaginary time integration. Note that a regularization parameter \(\varepsilon\), the lower limit of the integration in the $z$ coordinate,  has been included. 
        
        The action for the thermal AdS geometry is given by
        \begin{equation}\label{eq: explicit ads equation}
            {\cal I}^{(AdS)} = \frac{L^n}{\kappa^2}\beta_0 V_n\frac{1}{\varepsilon^{n+1}}\,,
        \end{equation}
        where \(\beta_0\) is related to \(\beta\) by requiring that both geometries have the same temporal period on the boundary, yielding
        \begin{equation}\label{eq: beta0 and beta}
            \beta_0  =  \sqrt{f(\varepsilon)}\;\beta\,.
        \end{equation}

        In order to remove the divergences as \(\varepsilon\to 0\), one can define the renormalized action as the difference
        \begin{equation}\label{eq: regularized scheme}
            \Delta \I \equiv \lim_{\varepsilon\to 0}\p{\I^{(BH)} -\I^{(AdS)}}.
        \end{equation}
        Substituting the horizon function \eqref{eq: schwarzschild horizon function} into relation \eqref{eq: beta0 and beta} and performing an expansion for small \(\varepsilon\), one finds the following expression for the renormalized action:
        \begin{equation}\label{eq: Sch explicit action}
            \Delta \I = - \frac{L^n}{2\kappa^2}\beta V_n\frac{1}{z_h^{n+1}}\,.
        \end{equation}

        Thermodynamic properties are derived from the partition function, \(Z\), which in the semiclassical approximation is related to the on-shell Euclidean action through   \cite{Hawking:1982dh}
        \begin{equation}\label{eq: partion function relation}
            \ln Z= - \Delta \I \,=\frac{L^n}{2 \kappa^2}\beta V_n\frac{1}{z_h^{n+1}}\,.
        \end{equation}
        
        From this relation, the internal energy is obtained as 
        \begin{equation}\label{eq: schw internal energy}
            \langle E \rangle = - \frac{\partial \ln Z}{\partial \beta} =  \frac{L^n}{2\kappa^2} V_n \frac{n}{z_h^{n+1}}\,,
        \end{equation}
        and the entropy as  

\begin{equation}\label{eq: schw entropy}
            S = \beta \langle E \rangle + \ln Z= \frac{L^n}{\kappa^2} V_n \frac{ 2 \pi }{z_h^n}\,.
        \end{equation}

        Thus, the off-shell free energy  \eqref{eq: off shell free energy definition} takes the form  
        \begin{equation}\label{eq: off shell free energy schwarzschild}
            {\cal F} = \frac{L^n}{2\kappa^2} V_n \p{\frac{n}{z_h^{n+1}} - \frac{4\pi}{z_h^n}\bar T}.
        \end{equation}

        Consequently, the \(z_h\)-component of the topological field \eqref{eq: topological field defition 2}  is given by
        \begin{equation}\label{eq: phi z_h schwarzschild}
            \phi^{_{z_h}} = \frac{\partial {\cal F}}{\partial z_h} = \frac{L^n}{2\kappa^2} V_n \frac{n}{z_h^{n+1}}\p{-\frac{(n+1)}{z_h} + 4\pi\bar T}\,. 
        \end{equation}

\begin{figure}[ht]
            \begin{subfigure}[h]{0.45\textwidth}
                \includegraphics[width=1\linewidth]{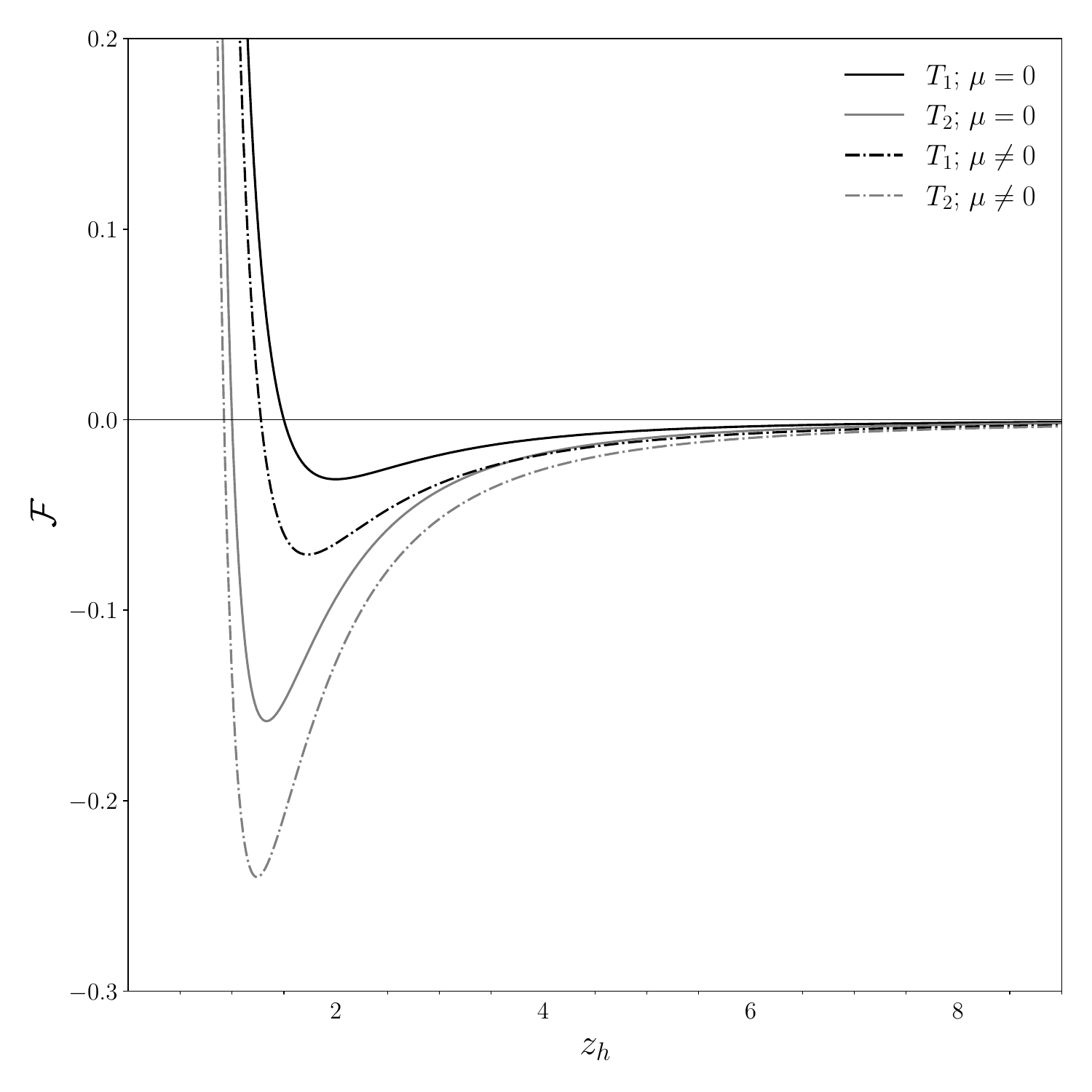}
                \caption{}
                \label{fig: off shell free energy}
            \end{subfigure}
            \hfill
            \begin{subfigure}[h]{0.5\textwidth}
                \includegraphics[width=.9\linewidth]{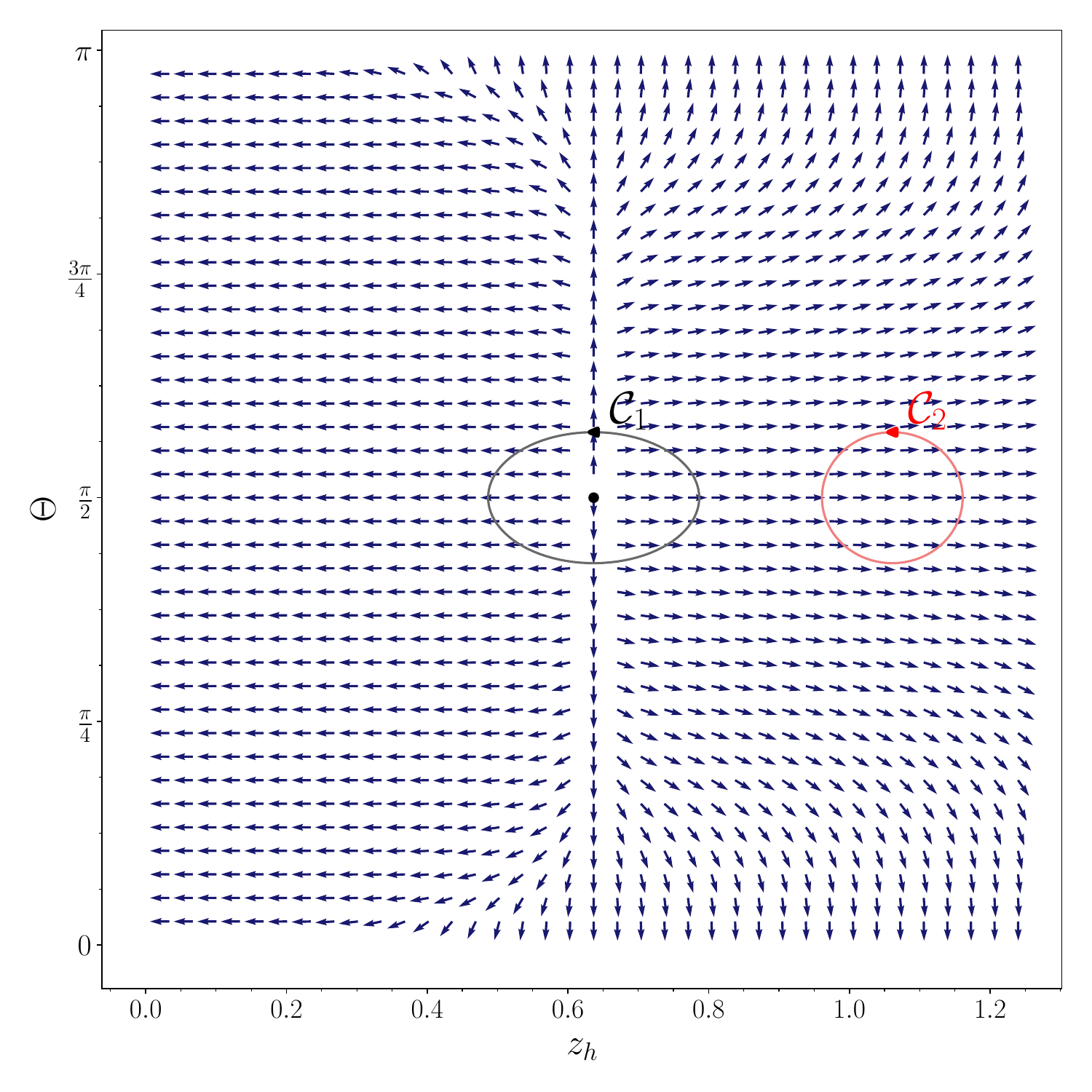}
                \caption{}
                \label{fig: topologic field bh schwarzschild}
            \end{subfigure}
            \caption{(a) Off-shell free energy for \(n=3\). The solid lines represent two different temperatures (\(\bar T_2 > \bar T_1\)) at vanishing chemical potential, while the dashed line corresponds to the same temperatures at finite chemical potential. (b) Normalized field \(\phi (z_h, \Theta)\). Contour \({\cal C}_1\) encloses a topological defect, in contrast to \({\cal C}_2\), which does not. In these plots, the constant \(L^3 V_3/\kappa^2\) was set to unity.}

        \end{figure}

        By definition, the roots of this field component correspond to extreme points of the off-shell free energy. In this case, there is only one root for finite values of \(z_h\), located at
        \begin{equation}\label{eq: horizon defect position}
            z_{hd} = \frac{\p{n+1}}{4\pi \bar T}\,.
        \end{equation}
        It is easy to verify that this point corresponds to a minimum of the off-shell free energy (as shown in Fig.~ \ref{fig: off shell free energy}). It is also straightforward to see that at this point the Hawking temperature \eqref{eq: Schwarzschild hawking temperature} is recovered. Thus, the vanishing point of the \(z_h\)-component and consequently the topological defect of the field \(\phi(z_h, \Theta)\) at \(\Theta=\pi/2\) (see Fig.~ \ref{fig: topologic field bh schwarzschild}) corresponds exactly to the black hole solution.

        In Fig.~ \ref{fig: topologic field bh schwarzschild}, the contour \({\cal C}_1\), which encloses the defect, is parametrized by the angle \(\vartheta\), utilizing the coordinates transformation \eqref{eq: coordenate transformation} with the identification \(x^1\equiv z_h\), \(x^2\equiv \Theta\) and setting \(B=\pi/2\) and \(A=z_{hd}\) as given by Eq. \eqref{eq: horizon defect position}. For the contour \({\cal C}_2\), we simply shift the value of \(A\) to exclude the defect.

\begin{figure}[ht]
            \begin{subfigure}[h]{0.45\textwidth}
                \includegraphics[width=1\linewidth]{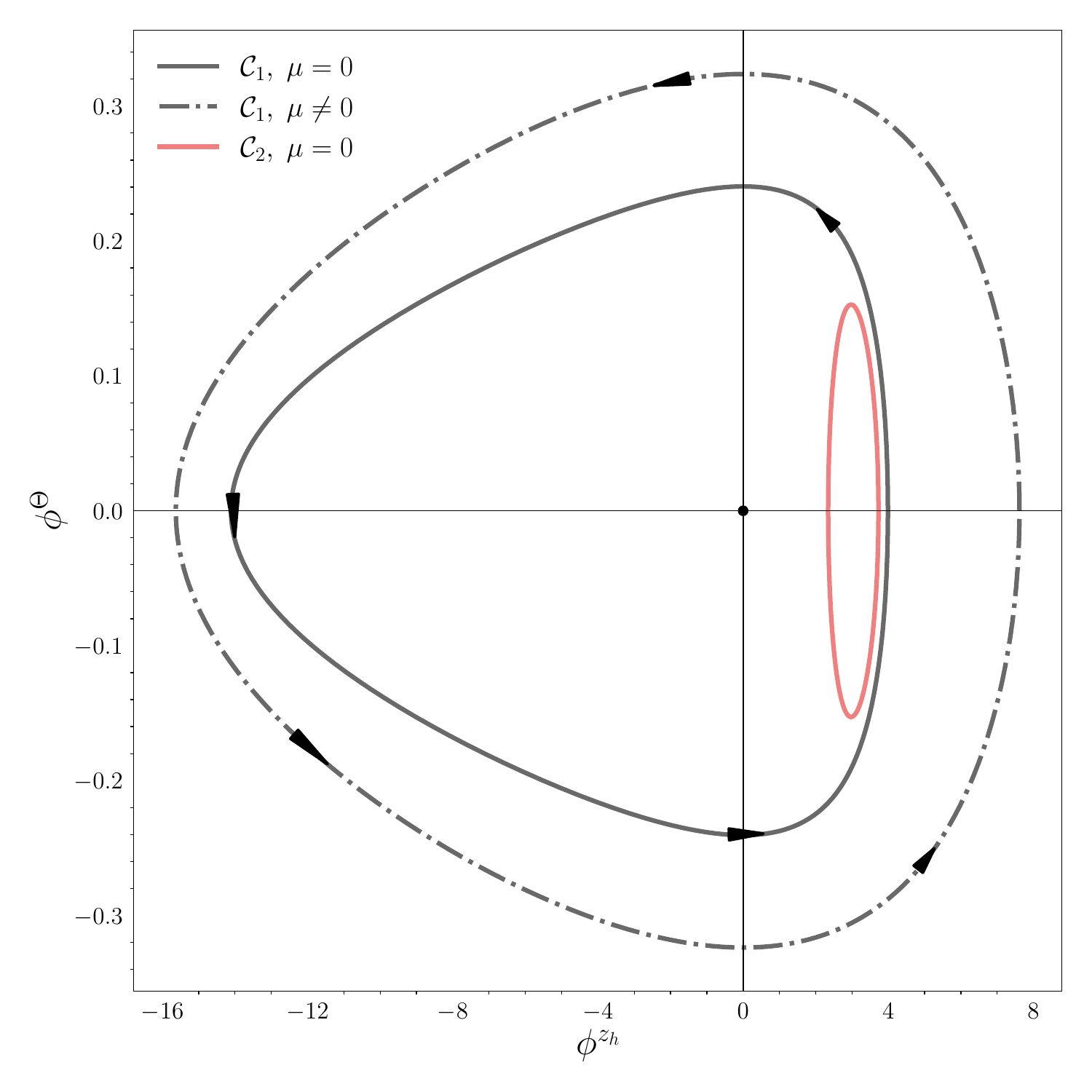}
                \caption{}
                \label{fig: vetor space}
            \end{subfigure}
            \hfill
            \begin{subfigure}[h]{0.5\textwidth}
                \includegraphics[width=.9\linewidth]{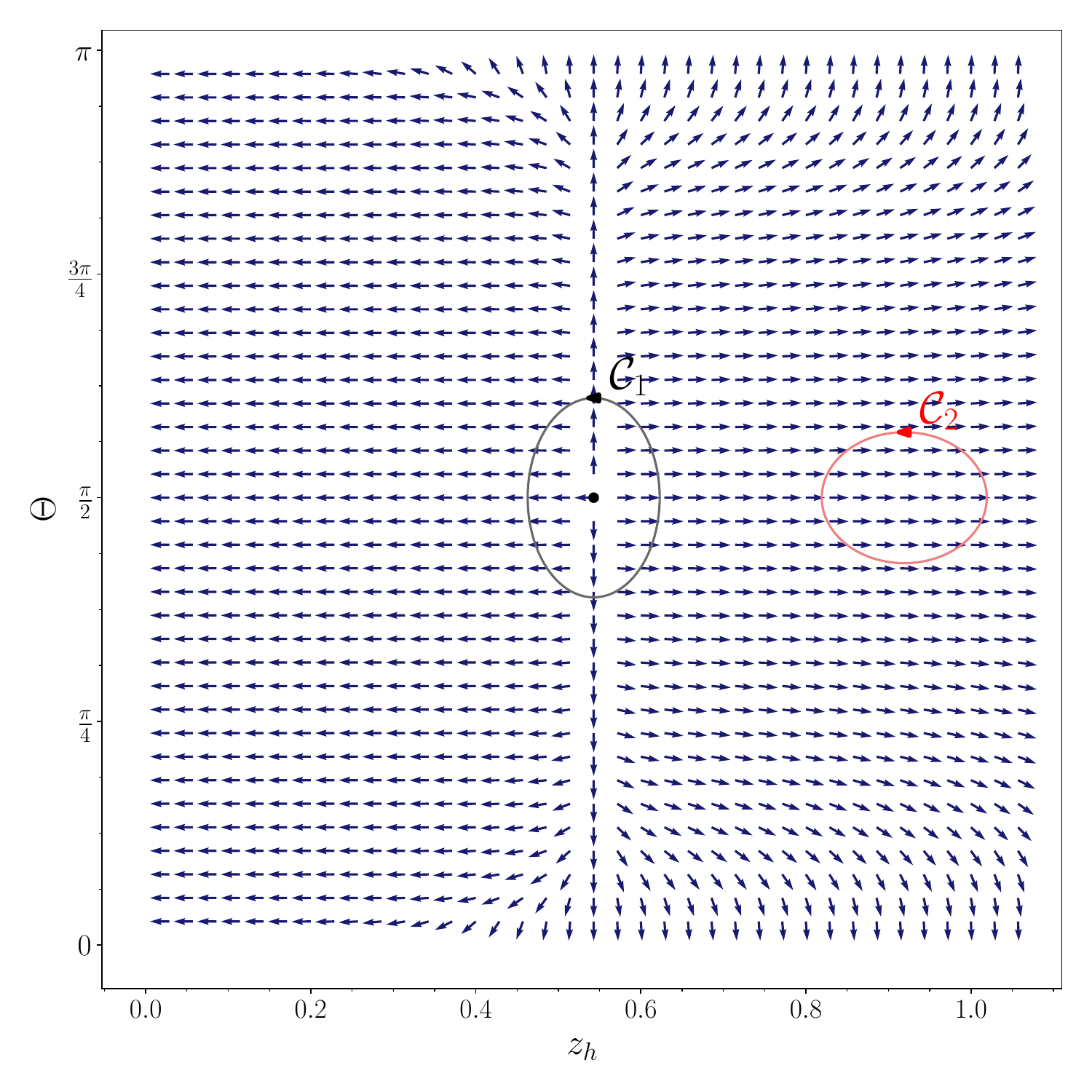}
                \caption{}
                \label{fig: topologic field bh reisser}
            \end{subfigure}
            \caption{
                    (a) Mapping the contours \({\cal C}\) in the vector space, the gray lines enclose the defect at the origin. Both Schwarzschild (solid line) and Reissner-Nordström (dash-dotted line) perform the loop in a counterclockwise direction, implying that the defect has \(w=+1\). The red line does not enclose the origin, indicating zero winding number. The red contour for the Reissner-Nordström was omitted for brevity.  (b) Normalized field obtained from \(\phi(z_h, \Theta)\) for the \(AdS_5\) Reissner-Nordström (\(n=3\)). Again, contour \({\cal C}_1\) encloses the topological defect but contour \({\cal C}_2\) does not. In these plots \(L^3 V_3/\kappa^2\) was set to unity.}
                
        \end{figure}

        Performing the integration in Eq. \eqref{eq: deflection angle integral} along contour \({\cal C}_2\), one finds that the winding number vanishes, as expected since no defect is enclosed. In contrast, the contour \({\cal C}_1\) gives \(w=+1\), meaning that the field performs a counterclockwise loop, Fig.~ \ref{fig: vetor space}. This result is typically associated with thermodynamically stable solutions \cite{Wei:2022dzw}, which is consistent with the fact that this point corresponds to the global minimum of the free energy \eqref{eq: off shell free energy schwarzschild}.

        Since there are no other defects in the bulk of the parameter space, the total topological charge is \(W=+1\). This is consistent with the analysis presented in \cite{Wei:2024gfz}, where this number is obtained by examining the field behavior at the boundary of the parameter space, a procedure shown to be equivalent via Eq.\eqref{eq: winding number as variation angle}. 

        Let us extend this analysis to the present case. By definition, the \(\Theta\)-component diverges in the horizontal edges 
        \begin{equation}\label{eq: phi component at edges}
            \phi^\Theta (\Theta=0)\to-\infty \quad \text{and} \quad\phi^\Theta \p{\Theta=\pi}\to +\infty
        \end{equation}
        meaning that for finite values of \(z_h\)-component, the vector field \(\phi\) points downward at the bottom edge and upward at the top edge. Consequently, the total topological charge, and thus the topological classification, depends solely on the behavior of the \(z_h\)-component at vertical edges.

        From Eq. \eqref{eq: phi z_h schwarzschild}, it is easy to see that the sign, and consequently the direction, of this component depends on the Hawking temperature term inside the parentheses. In the limit \(z_h\to0\) (high temperature limit), this term becomes dominant, leading \(\phi^{z_h}\to-\infty\). Since this divergence scales as \(z_h^{-(n+2)}\), it dominates over the divergence of the \( \phi^\Theta\) and forces the vector field  \(\phi\) to point leftward along the entire left edge of the parameter space.

        In the zero temperature limit (\(z_h\to\infty\)), the term proportional to \(\bar T\) dominates, and thus \(\phi^{z_h}\) points rightward. However, it should be noted that the \(\phi^{z_h}\) decays rapidly  to zero; consequently, the orientation of the vector field \(\phi\) is determinate by the \(\phi^{\Theta}\)-component. As previously discussed, the vector points downward at the bottom corner and upward at the top corner. Therefore, following the path along the boundary, the field performs an anticlockwise loop as expected.  It is interesting to note that edge at \(z_h\to\infty\), contributes with \(+1/2\) to total winding number.

        Within the four topological classes presented in \cite{Wei:2024gfz}, this black hole belongs to the \(W^{1+}\) class.  Note, however, that in contrast to geometries with compact boundaries, there is only one solution here, which remains stable at both low and high temperatures.

    \subsection{AdS Reissner–Nordström}

        Under the gauge/gravity duality, a medium with a finite chemical potential on the gauge theory side is represented by the charged black hole, the Reissner-Nordström geometry \cite{Braga:2019xwl, Braga:2024nnj, Braga:2025wlx, Colangelo:2010pe, Colangelo:2012jy}. In this case, the horizon function in the metric \eqref{eq: bh ads D metric} is given by
        \begin{equation}\label{eq: RN horizon function}
            f(z) = 1 - \p{1+q^2 z_h^{2n}}\frac{z^{n+1}}{z_h^{n+1}} + q^2 z^{2n}\,,
        \end{equation}
        where \(q\) is a parameter proportional to the black hole charge. 
        
        By solving the Einstein-Maxwell equations, one finds that this charge generates an electric field derived from a gauge potential in the form
        \begin{equation}\label{eq: vector potential}
            A_t(z) = \mu  - \frac{a}{(n-1)} z^{n-1}\,.
        \end{equation}
        Here, \(a\) is a constant related to \(q\) by 
        \begin{equation}\label{eq: q as a}
            q^2 = \frac{\kappa^2}{\p{n-1}n g^{2}_{{n+2}}L^2}a^2
        \end{equation}
        being \(g_{n+2}\) the gauge coupling constant. The boundary value of the gauge field \(\p{z\to0}\) corresponds to the chemical potential \(\mu\). By requiring the gauge potential to vanish at the horizon, \(A_t(z_h)=0\) \cite{Lee:2009bya}, we establish a relation between the chemical potential and the parameter \(q\) as
        \begin{equation}\label{eq: relation between q and mu}
            q = \frac{\bar \mu}{z_h^{n-1}} \quad\text{with}\quad \bar \mu \equiv \sqrt{\frac{\p{n-1}\kappa^2}{n g^{2}_{n+2} L^2}}\mu\,.
        \end{equation}
       
        The Hawking temperature, derived from the horizon function \eqref{eq: RN horizon function}, is given by
        \begin{equation}\label{eq: RN hawking temperature for q}
            T  = \frac{1}{4\pi z_h}\left [\p{n+1}-\p{n-1}q^2 z_h^{2 n}\right]\,.
        \end{equation}
        Alternatively, applying relation \eqref{eq: relation between q and mu}, the temperature can be expressed in terms of the chemical potential as
        \begin{equation}
            T = \frac{1}{4\pi z_h}\left [\p{n+1}-\p{n-1}\bar \mu^2z_h^2\right]\,.
        \end{equation}

        The action for this space now includes the Maxwell term and can be expressed as
        \begin{align}
            {\cal I}&={\cal I}_{EH}+{\cal I}_M\nonumber\\
             &= - \frac{1}{2 \kappa^2}\int d^{n+2}x\sqrt{|g|}\p{{\cal R}- \Lambda} + \frac{1}{4 g_{{n+2}}^2}\int d^{n+2}x \sqrt{|g|}F_{\sigma \rho}F^{\sigma\rho}\label{eq: RN AdS action}
        \end{align} 
        where \(F_{MN}=\partial_M A_N - \partial_N A_M\) is the strength tensor. From the vector potential in Eq. \eqref{eq: vector potential}, it is straightforward to show that the only non-vanishing components are  \(F_{tz}=-F_{zt}= a z^{n-2}\). Consequently, the Maxwell part of the action reduces to
        \begin{equation}\label{eq: Maxwell action}
            {\cal I}_M = -\frac{V_n L^n}{2\kappa^2}\beta n q^2 z_h^{n-1}\,.
        \end{equation}

        For the gravitational part of the action, it is necessary to account for the correction to the Ricci scalar due to the presence of the electric field. To do so, let us consider Einstein's equations 
        \begin{equation}\label{eq: Einstein Equation}
            {\cal R}_{\mu\nu} -\frac{1}{2}g_{\mu\nu}{\cal R}+\frac{1}{2}g_{\mu\nu}\Lambda = \kappa^2   {\cal T}_{\mu\nu}\,,
        \end{equation}
        where \({\cal T}_{\mu\nu}\) is the energy-momentum tensor. For a Maxwell field reads 
        \begin{equation}\label{eq: energy-momentum tensor}
            {\cal T}_{\mu\nu} = \frac{1}{g_{n+2}}\p{F_{\mu\rho}F^{\;\;\rho}_\nu-\frac{1}{4}g_{\mu\nu}F^{\sigma \rho}F_{\sigma \rho}}\,.
        \end{equation}

        Taking the trace of Eq. \eqref{eq: Einstein Equation} yields
        \begin{equation}
            {\cal R} = \frac{\p{n+2}}{n}\Lambda - \frac{2\kappa^2}{n}{\cal T}
        \end{equation}
        with the trace of the energy-momentum tensor given by 
        \begin{equation}
            {\cal T} \equiv g^{\mu\nu}{\cal T}_{\mu\nu} = -\frac{\p{n-2}}{4g_{n+2}}F^{\sigma \rho}F_{\sigma \rho}\,.
        \end{equation}
        By using the strength tensor \(F_{tz} = a z^{n-2}\) and substituting these expressions into the gravitational part of the action \eqref{eq: RN AdS action} combined with the cosmological constant from \eqref{eq: Ricci and cosmological constant}, one obtains
        \begin{equation}
            {\cal I}_{EH} = \frac{V_n L^n}{\kappa^2}\beta\p{\frac{1}{\varepsilon^{n+1}} - \frac{1}{z_h^{n+1}}}+\frac{V_n L^n}{2\kappa^2}\beta\p{n-2}q^2 z_h^{n-1}\,,
        \end{equation}
        where the relation \eqref{eq: q as a} has been applied. Consequently, the total action for the RN-AdS geometry is 
        \begin{equation}
            {\cal I}^{(BH)} = \frac{V_n L^n}{\kappa^2}\beta\p{\frac{1}{\varepsilon^{n+1}} - \frac{1}{z_h^{n+1}}-q^2 z_h^{n-1}}\,.
        \end{equation}

        The background AdS geometry (in the limit \(z_h\to \infty\) and \(q\to0\)) is still given by Eq. \eqref{eq: explicit ads equation}. Thus, after inserting the horizon function \eqref{eq: RN horizon function} into relation \eqref{eq: beta0 and beta}, performing an expansion for small \(\varepsilon\) and  employing the subtraction scheme \eqref{eq: regularized scheme}, the regularized action is obtained as
        \begin{equation}\label{eq: RN explicit action}
            \Delta{\cal I} = -\frac{V_n L^n}{\kappa^2}\beta\p{\frac{1}{z_h^{n+1}}+ q^2 z_h^{n-1}}\,.
        \end{equation}

        Note that the presence of the charge lowers the value of the action, making the charged geometry always thermodynamically favored over the neutral (see Fig.~ \ref{fig: off shell free energy}). By establishing the connection with thermodynamics in the grand-canonical ensemble through the semi-classical approximation in Eq. \eqref{eq: partion function relation}, the internal energy
        \begin{equation}
            \langle E\rangle = n\frac{L^n}{2\kappa^2}V_n \p{\frac{1}{z_h^{n+1}}+\frac{\bar \mu^2}{z_h^{n-1}}}\,,
        \end{equation}
        the entropy
        \begin{equation}
            S = \frac{ L^n}{\kappa^2}V_n \frac{2 \pi}{z_h^n}\,,
        \end{equation}
        and the total charge 
        \begin{equation}
            Q =  n\frac{L^n}{\kappa^2}V_n \frac{\bar \mu}{z_h^{n-1}}\,= n\frac{L^n}{\kappa^2}V_n q\,,
        \end{equation}
        can be readily obtained.

        By constructing the off-shell free energy as 
        \begin{align}
            {\cal F} &= \langle E\rangle - \bar T S -\bar \mu Q\nonumber\\
            &= \frac{V_n L^n}{2\kappa^2}\p{\frac{n}{z_h^{n+1}} - n\frac{\bar \mu^2}{z_h^{n-1}}-4\pi\frac{\bar T}{z_h^n}}\,,\label{eq: RN off shell free energy}
        \end{align}
        consequently, the \(z_h\)-component of the topological field for this case is expressed as 
        \begin{equation}\label{eq: RN phi component}
            \phi^{z_h} = \frac{L^n}{2\kappa^2} V_n \frac{n}{z_h^{n+1}}\p{-\frac{(n+1)}{z_h} + \p{n-1}\bar \mu^2 z_h + 4\pi\bar T}\,.
        \end{equation}

        It is evident that this component vanishes when the Hawking temperature is recovered. For positive finite values of \(z_h\), this occurs at point 
        \begin{equation}\label{eq: RN defect location}
            z_{hd} = \frac{-4\pi \bar T+\sqrt{\p{4\pi \bar T}^2+4\p{n^2-1}\bar\mu^2}}{2\p{n-1}\bar \mu^2}\,.
        \end{equation}
        This location also corresponds to the stable minimum of the off-shell free energy, see Fig.~\ref{fig: off shell free energy}.

        Since the \(\Theta\)-component remains unchanged, the defect is located at \(\Theta=\pi/2\) and \(z_{hd}\) as given by the equation \eqref{eq: RN defect location}, as shown in figure \ref{fig: topologic field bh reisser}. As in the previous case, we enclose the defect with a contour \({\cal C}_1\) parametrized by the coordinates \eqref{eq: coordenate transformation}. By performing the integral \eqref{eq: deflection angle integral}, a winding number \(w=+1\) is found, meaning that the field rotated in an anticlockwise direction along the contour \({\cal C}_1\). This matches the Schwarzschild case, as shown in Fig.~\ref{fig: vetor space}, notably, the primary difference between the two cases lies in the vector intensity.

        There is only one defect in the bulk, resulting in a total topological charge of \(W=+1\). This is confirmed by the behavior of the vector field at the boundary, which, as easily verified from \eqref{eq: RN phi component}, remains identical to that of the neutral case. Consequently, the RN-AdS geometry with non-compact boundary belongs to the same topological class as the Schwarzschild solution, \(W^{1+}\), with the presence of charge serving only to shift the defect position.


\section{\label{sec: Sec3} Topological class of the phenomenological \(AdS/QCD\) model}

    As previously mentioned, the Hawking-Page transition in the gravity dual corresponds to the confinement/deconfinement transition in the gauge theory. In the \(AdS/QCD\) context, this represents a phase transition from a hadronic confined phase to a plasma of quarks and gluons (deconfined phase).   

    However, both actions \eqref{eq: Sch explicit action} and \eqref{eq: RN explicit action} vanish only in the limit \(zh\to\infty\), which corresponds to the zero-temperature limit. This implies that the stable minimum of the off-shell free energies, given in Eqs.\eqref{eq: off shell free energy schwarzschild} and \eqref{eq: RN off shell free energy}, is always the black hole solution, for any non-vanishing temperature.

    This is a consequence of the fact that the geometry is conformal, not possessing any energy scale. As a consequence, the plasma phase exists at all temperatures, which is in disagreement with the known properties of \(QCD\). In order to obtain a holographic representation that reproduces a thermal transition like in the QGP, it is necessary to introduce an energy scale through some \(AdS/QCD\) model.  Since they are defined in five dimensions, we will focus on the case \(n=3\) in the following analysis.

    \subsection{Overview of the HP transitions in  a phenomenological model}

        \begin{figure}[ht]
            \begin{subfigure}[h]{0.45\textwidth}
                \includegraphics[width=1\linewidth]{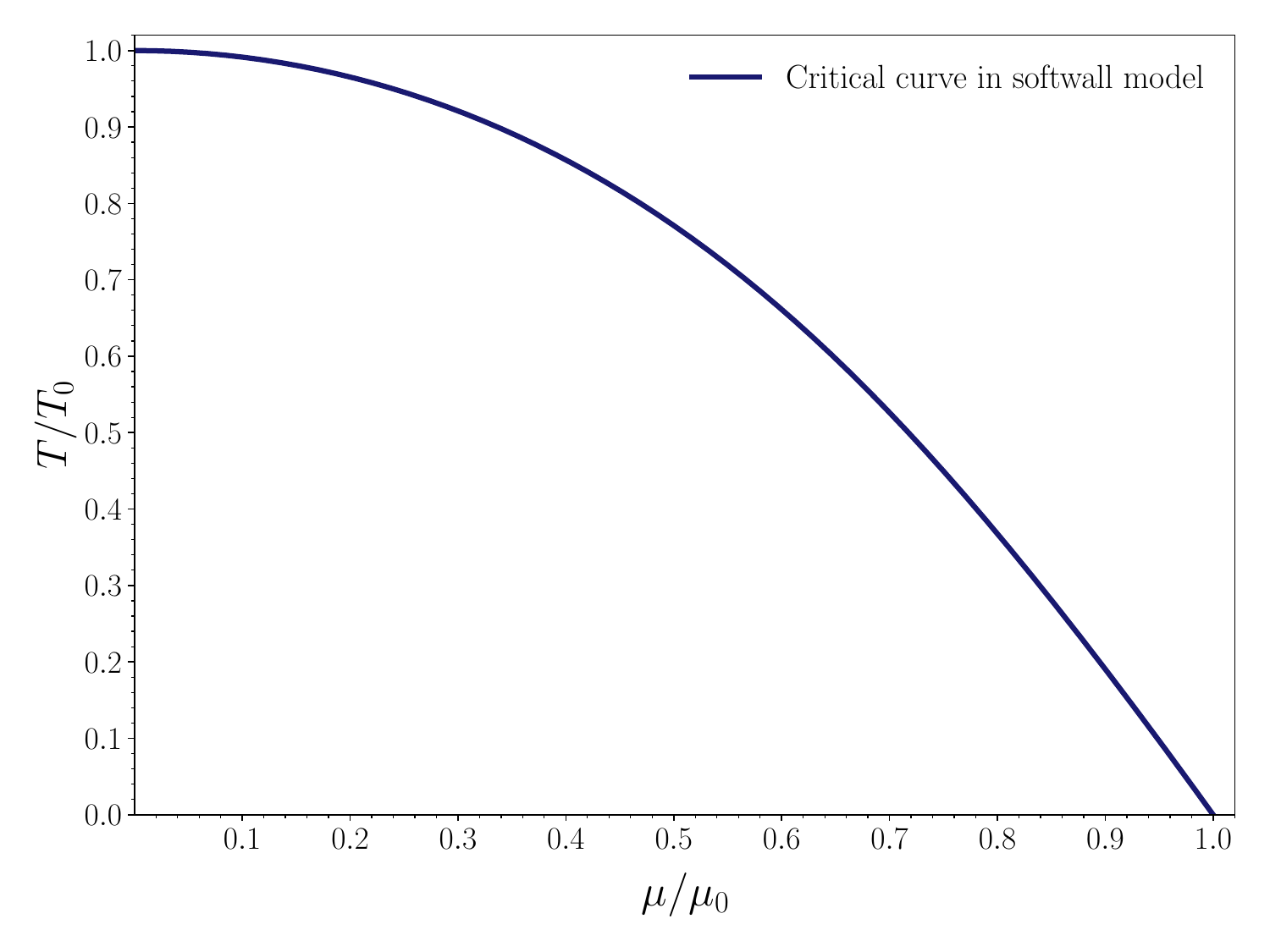}
                \caption{}
                \label{fig: dissociation curve}
            \end{subfigure}
            \hfill
            \begin{subfigure}[h]{0.45\textwidth}
                \includegraphics[width=1.\linewidth]{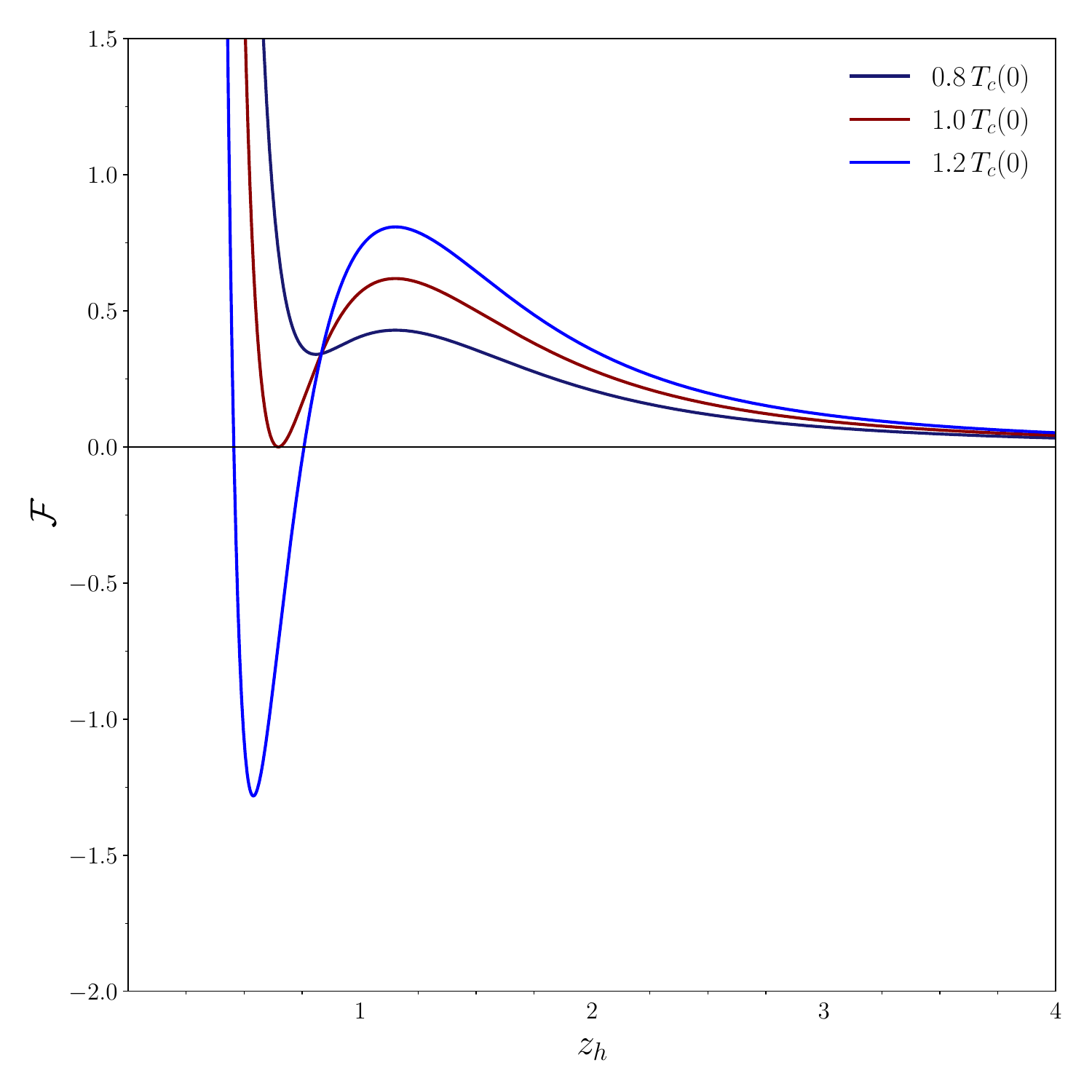}
                \caption{}
                \label{fig: off shell free energy for sw}
            \end{subfigure}
            \caption{
                          (a) Phase diagram for the deconfinement transition for the  SW model. (b) Off-shell free energy for the SW model at increasing temperatures with \(\bar \mu=0\) in units of \(T_c(0)\). The red line corresponds to the HP transition, where the minima, \(z_{hd}\) and \(z_h\to\infty\), coincide at \({\cal F} =0\). For these plots, \(L^3 V_3/\kappa^2\) was set to unity. 
                } 
            
        \end{figure}

     Here, we analyze a model in which an energy scale is introduced through a modification of the gravitational action, without considering backreactions. Specifically, we consider the Soft-Wall (SW) model \cite{Karch:2006pv}, defined by introducing a dilaton field of the form \(\Phi(z) = c^2z^2\)  into action \eqref{eq: RN AdS action}, where \(c\) is the  parameter that represents an energy scale. The action is now expressed by 
        \begin{equation}\label{eq: sw action}
            {\cal I}=- \frac{1}{2 \kappa^2}\int d^{5}x\sqrt{|g|}\,e^{-c^2z^2}\p{{\cal R}- \Lambda} + \frac{1}{4 g_{{5}}^2}\int d^{5}x \sqrt{|g|}\,e^{-c^2z^2}F_{\sigma \rho}F^{\sigma\rho}\,.
        \end{equation}
                
        Following the same procedure as in the previous section, the regularized action for the SW model is found to be
        \begin{align}   
            \Delta{\cal I} =\frac{VL^3}{\kappa^2}\beta\frac{1}{z_h^4}\left(e^{-c^2 z_h^2}\left[-1+c^2 z_h^2\right]+\frac{1}{2}+c^4 z_h^4\text{Ei}(-c^2 z_h^2)\right.\nonumber\\
            \left.+\bar \mu^2 z_h^2\left[\frac{e^{-c^2 z_h^2}}{c^2 z_h^2}-\frac{1}{c^2 z_h^2}+\frac{1}{2}\right]\right)\label{eq: sw explicit action}
        \end{align} 
        where the exponential integral is defined as \(\text{Ei}(x)=-\int_{-x}^\infty\tfrac{e^{-t}}{t}dt\). Note that in the limit \(c\to 0\) the Reissner-Nordström action \eqref{eq: RN explicit action} is recovered.   The determination of the critical temperature where the action vanishes is performed numerically. The phase diagram obtained, which exhibits the dependence on the chemical potential, is shown in Fig.~\ref{fig: dissociation curve}.

        Performing a numerical analysis for the case \(\bar\mu=0\) the standard result  \(T_c(0)=0.4917\, c\) from Ref. \cite{Herzog:2006ra} is recovered. In the high chemical potential limit, we find that the critical chemical potential depends on \(c\) as \(\bar \mu_c = 1.42832\, c\). This result differs from the one obtained in Ref.\cite{Braga:2024nnj}, as those authors consider a charged AdS space. 

        Phenomenologically, the parameter \(c\) can be determined by the best fit for the experimental data for the masses of  \(\rho\) meson states.   This value is usually set as \(c=388 \; MeV\) \cite{Herzog:2006ra}~. This choice leads to a critical chemical potential \(\mu_c =358.12\; MeV\), where the constant  in the relation \eqref{eq: relation between q and mu} for \(AdS_5\)  is \(\sqrt{5/12}\) (see \cite{Braga:2025wlx} for details). Expressing this in terms of the baryon chemical potential, our prediction becomes \(\mu_{bc}=3\mu_c=1074.361\; MeV\). This is a reasonable result, as lattice \(QCD\) predictions for this critical point lie around \(1200 \;MeV\) \cite{Braga:2024nnj,Sarkar:2010zza, Colangelo:2010pe}.

    \subsection{Off-shell Free energy behavior}
        
        Let us examine how the introduction of an energy scale affects the topological classification established in Sec\ref{sec: Sec2} for pure AdS geometries. First, we consider the case where \(\bar\mu=0\), which corresponds to the Schwarzschild limit. From the action in Eq.\eqref{eq: sw explicit action}, the internal energy is obtained as 
        \begin{equation}\label{eq: sw internal energy}
            \langle E \rangle = \frac{L^3 V}{\kappa^2}\frac{1}{z_h^4}\p{e^{-c^2z_h^2}\p{3+c^2z_h^2}-\frac{3}{2}+c^4z_h^4 \text{Ei}\p{-c^2z_h^2}},
        \end{equation}
        and the entropy is given by
        \begin{equation}\label{eq: sw entropy}
            S = \frac{L^3 V}{\kappa^2}\frac{\pi}{z_h^3}\p{4e^{-c^2z_h^2} -2}.
        \end{equation}
        Note that in the limit \(c\to 0\), one recovers the Schwarzschild internal energy and entropy from Eqs. \eqref{eq: schw internal energy} and \eqref{eq: schw entropy}. 

        Using the off-shell free energy definition from Eq.\eqref{eq: off shell free energy definition}, the following expression is obtained for the SW model
        \begin{equation}\label{eq: sw off shell free energy}
            {\cal F} =\frac{L^3V}{\kappa^2}\frac{1}{z_h^4}\left[{e^{-c^2z_h^2}\p{3+c^2z_h^2}-\frac{3}{2}+c^4z_h^4 \operatorname{Ei}\p{-c^2z_h^2}}-\pi z_h\bar T\p{4e^{-c^2z_h^2}-2}\right].
        \end{equation}

        \begin{figure}[ht]
            \begin{subfigure}[h]{0.45\textwidth}
                \includegraphics[width=1\linewidth]{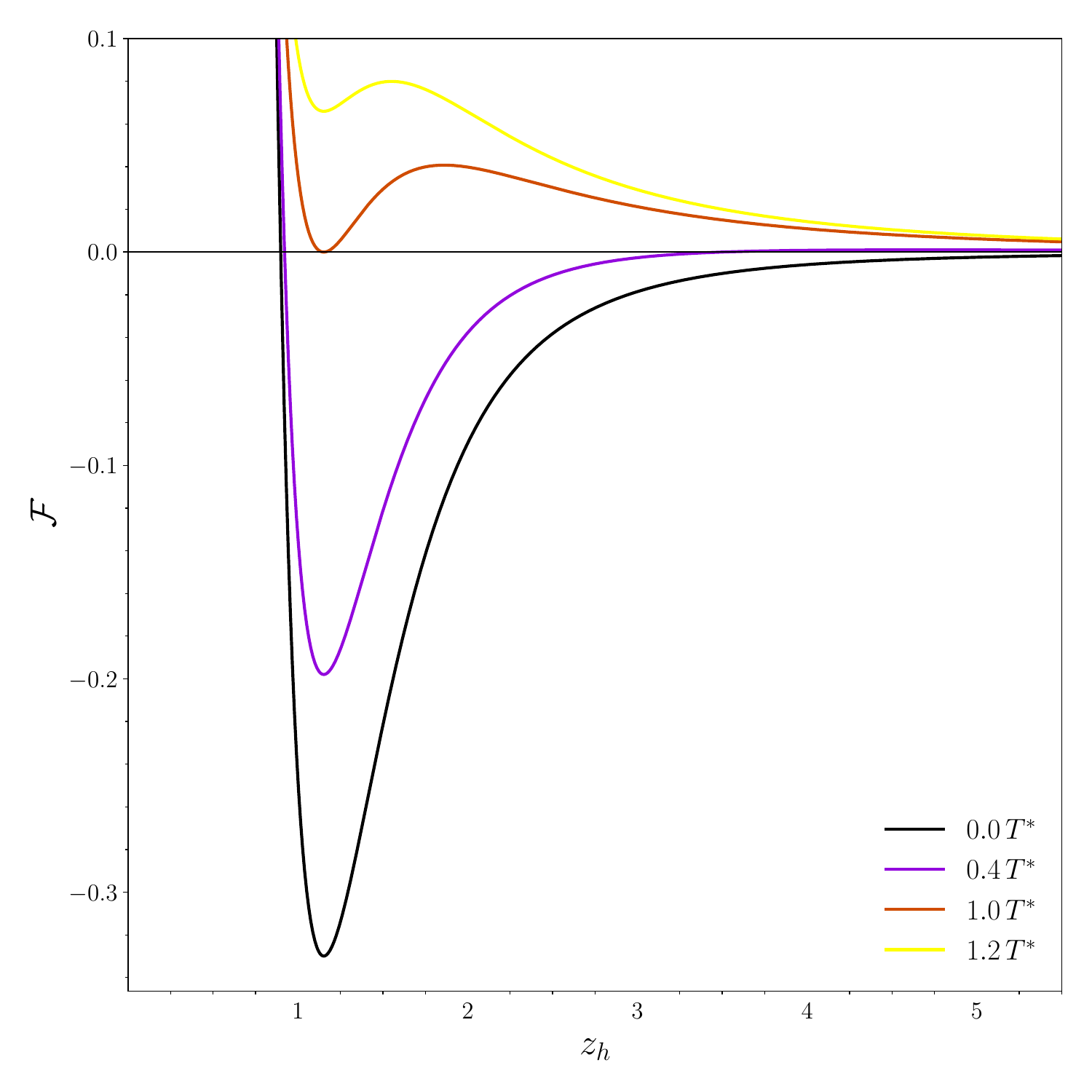}
                \caption{}
                \label{fig: sw sch low temperature behavior}
            \end{subfigure}
            \hfill
            \begin{subfigure}[h]{0.45\textwidth}
                \includegraphics[width=1.\linewidth]{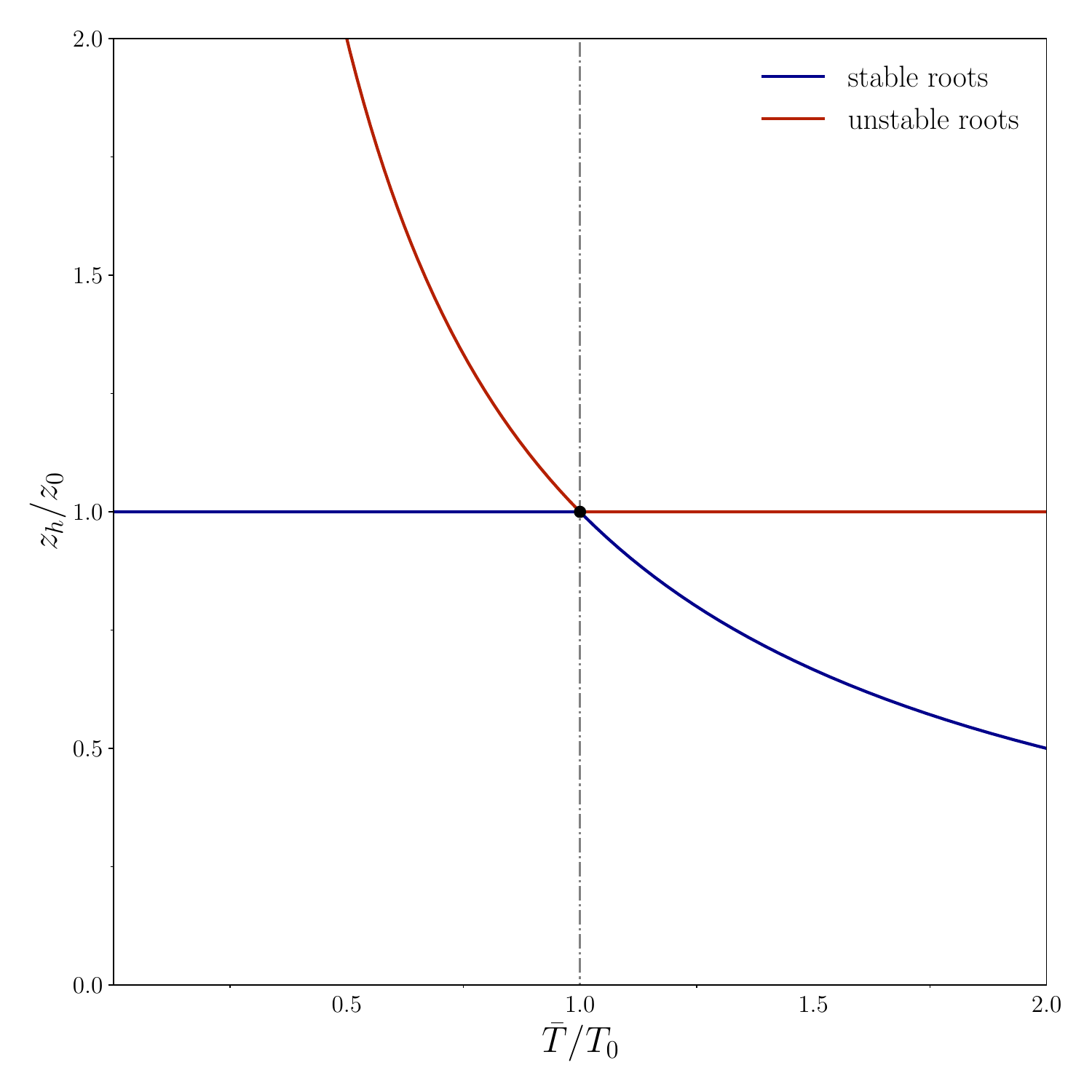}
                \caption{}
                \label{fig: plot roots of phi}
            \end{subfigure}
            \caption{
                    (a) Free energy at low temperatures. As the temperature increases, the minimum shifts, and a transition occurs when the minimum reaches zero, indicating the end of the confined phase. (b) Temperature dependence of the \(\phi^{z_h}\) roots. To the left of the gray dashed line, \(z_0\) is the stable root, while to the right, \(z_{hd}\) becomes stable; the black dot marks the transition point.
               }
                
        \end{figure}
        
        Consequently, the \(z_h\)-component of the vector field reads 
        \begin{equation}\label{eq: sw phi zh}
            \phi^{z_h} = \frac{L^3V}{\kappa^2}\frac{2}{z_h^5}e^{-c^2z_h^2}\p{-1+\pi \bar T z_h}\p{6+4c^2z_h^2-3e^{c^2z_h^2}}.
        \end{equation}
        By definition, the roots of \(\phi^{z_h}\) are the extrema of the free energy, representing stable or unstable states of the system. From the two factors inside parentheses in Eq. \eqref{eq: sw phi zh} we find two finite, real, and positive solutions. The first is
        \begin{equation}\label{eq: hawking defect}
            z_{hd} = \frac{1}{\pi \bar T}\,,
        \end{equation}
               which represents the black hole solution, the point where the Hawking temperature is recovered. The second root is given by
        \begin{equation}\label{eq: z0 minimum}
            z_{0} = \frac{1}{c}\sqrt{\frac{1}{2}\p{-3-2k}} \quad\text{with} \quad k ={\cal W}_{-1}(-\tfrac{3}{4}e^{-3/2})
        \end{equation}
        where \({\cal W}_k(z)\) is the Lambert \(\cal W\) function (or product logarithm). This root represents a state related to the confinement energy scale, as evidenced by the \(1/c\) factor. Also note that in the limit \(c\to 0\), this point coincides with the root at \(z_h\to\infty\).

        The dynamics of the system is described from the free energy as follows: the minima correspond to stable states, while the maxima represent unstable states. The global minimum determines the state of the system. At any temperature where the free energy values of the two minima coincide, which in this case occurs at \({\cal F}=0\), the system undergoes a first-order phase transition, as illustrated in Fig.~\ref{fig: off shell free energy for sw} for the HP transition discussed at the beginning of this section.

        Note that, in contrast to the pure AdS geometry case, taking \(\bar T = 0 \) in Eq.\eqref{eq: sw phi zh} reveals that \(z_0\) still exists, as it is independent of temperature. Inserting this root back into Eq.\eqref{eq: sw off shell free energy} yields a negative free energy for this minimum, as shown in the Fig.~\ref{fig: sw sch low temperature behavior}. Consequently, apparently  this represents a global minimum.  However, as the temperature increases, the free energy of the \(z_0\) minimum increases, eventually \({\cal F}=0\) at a temperature \(T^*\) in   an apparent global transition, which we will analyze in detail in the next section. The fact that \(\cal F\) increases with temperature at this minimum implies that the state possesses a negative entropy, rendering it unphysical. Thus, the only true physical minimum of the system for temperatures below \(T_c\p{0}\) is the boundary limit \(z_h\to\infty\), the thermal AdS, or in the dual theory, the confined hadronic state.

        \afterpage{
    \clearpage
    \begin{figure}[p]
        \centering
        \begin{subfigure}[t]{.45\linewidth}
            \includegraphics[width=\linewidth, height =.24\textheight]{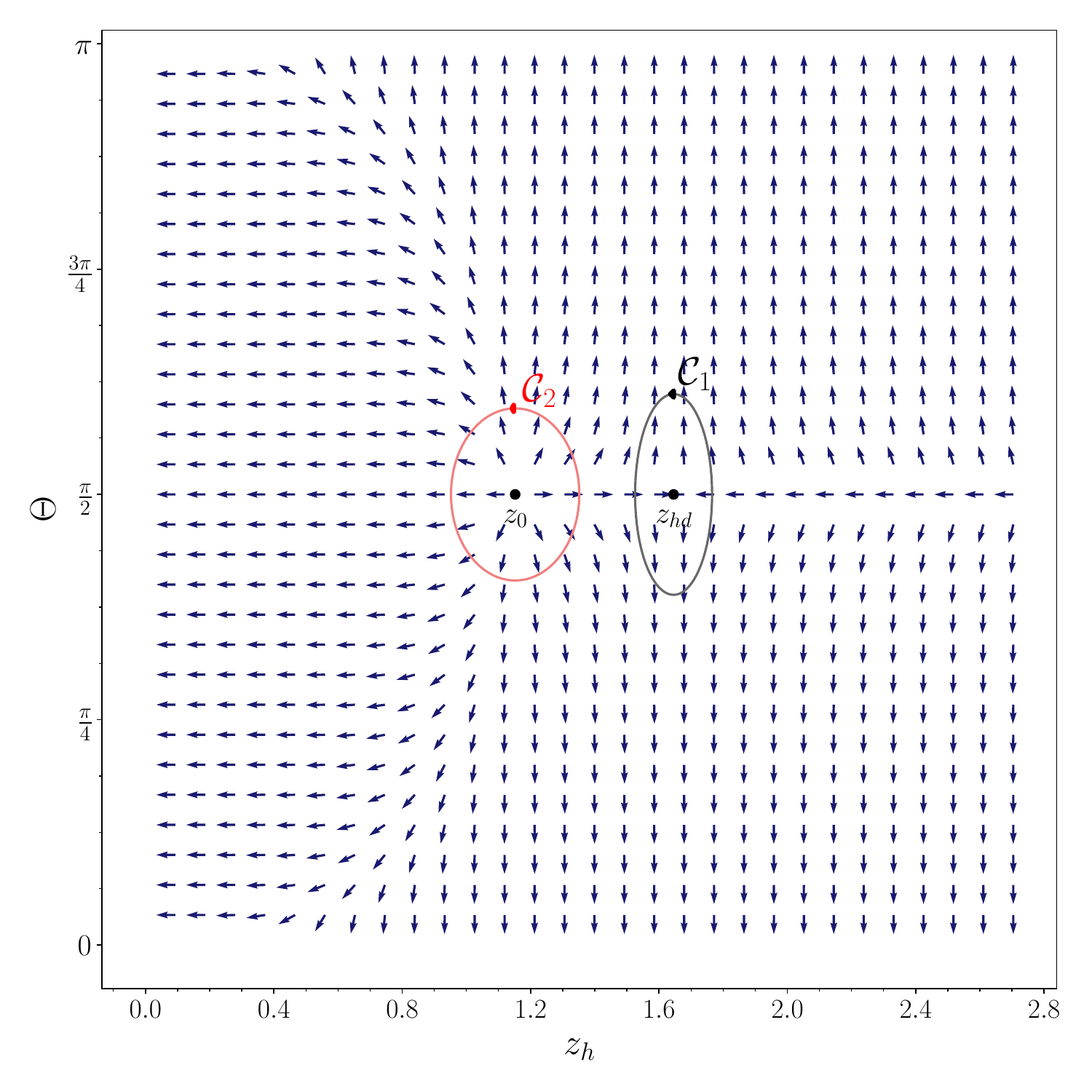}
            \caption{\(T < T_0\)}
            \label{fig: sw vector field below T0}
        \end{subfigure}\qquad
        \begin{subfigure}[t]{.45\linewidth}
            \includegraphics[width=\linewidth, height =.24\textheight]{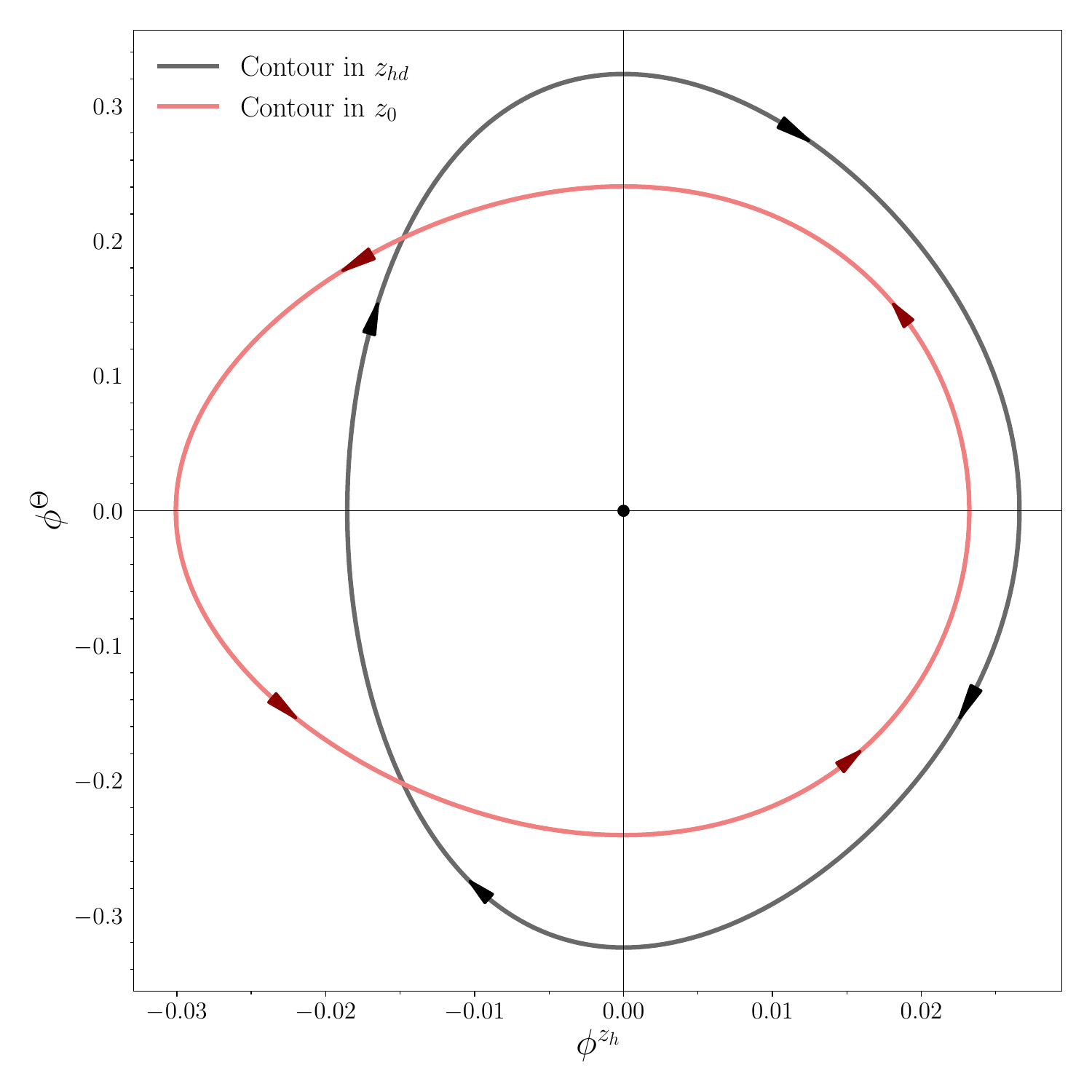}
            \caption{\(T < T_0\)}
            \label{fig: sw vector space below T0}
        \end{subfigure}\\
        \begin{subfigure}[t]{.45\linewidth}
            \includegraphics[width=\linewidth, height =.24\textheight]{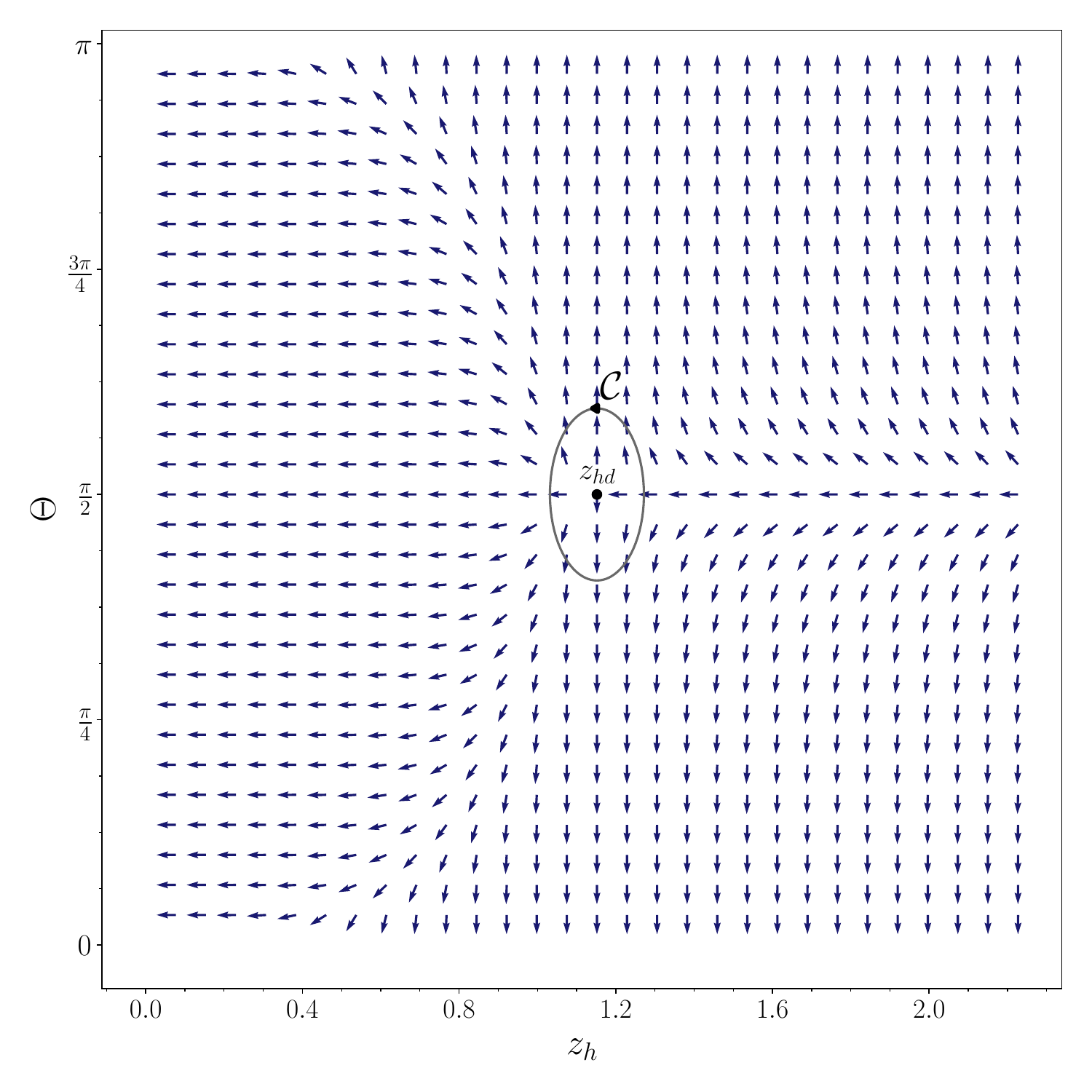}
            \caption{\(T = T_0\)}
            \label{fig: sw vector field at T0}
        \end{subfigure}\qquad
        \begin{subfigure}[t]{.45\linewidth}
            \includegraphics[width=\linewidth, height =.24\textheight]{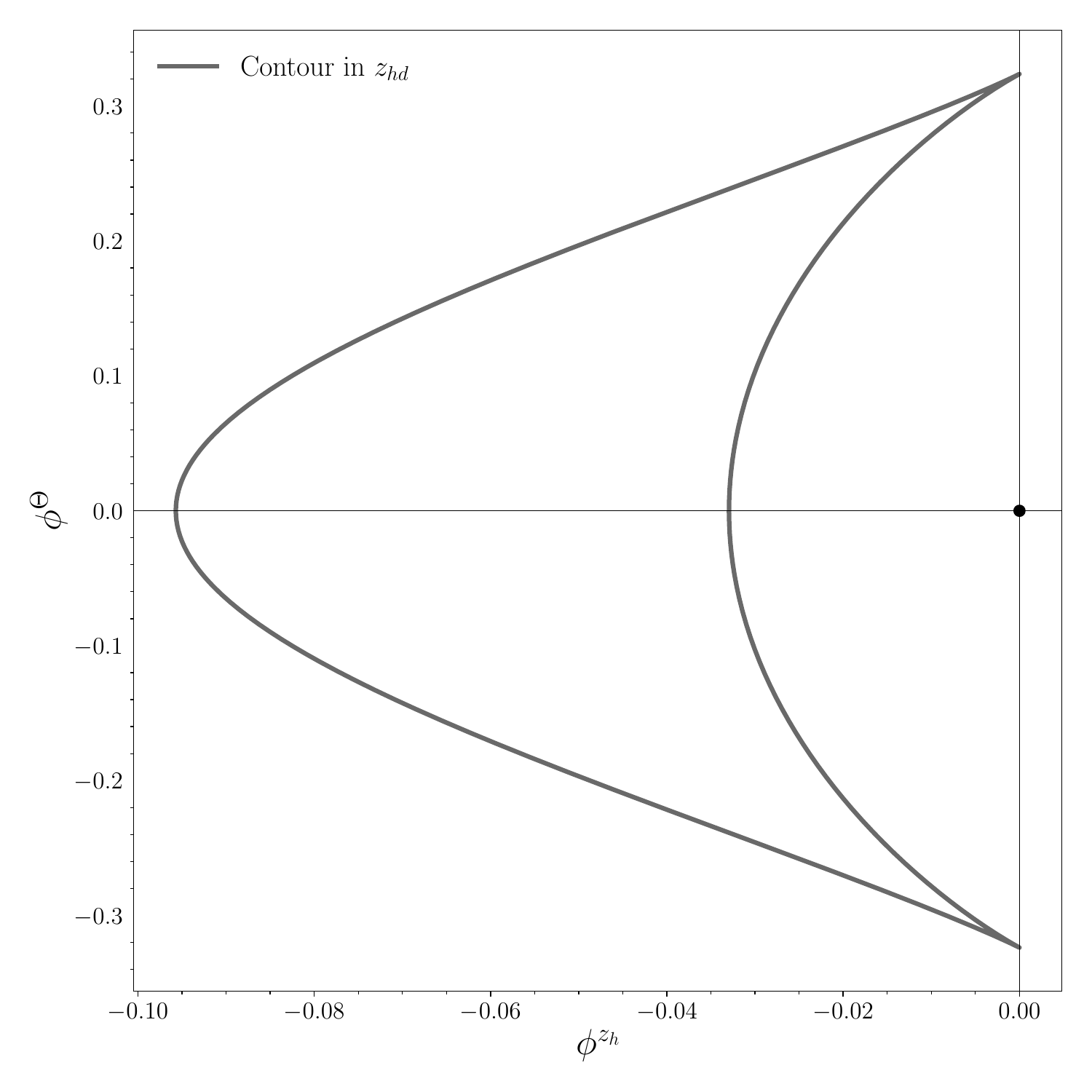}
            \caption{\(T = T_0\)}
            \label{fig: sw vector space at T0}
        \end{subfigure}\\
        \begin{subfigure}[t]{.45\linewidth}
            \includegraphics[width=\linewidth, height =.24\textheight]{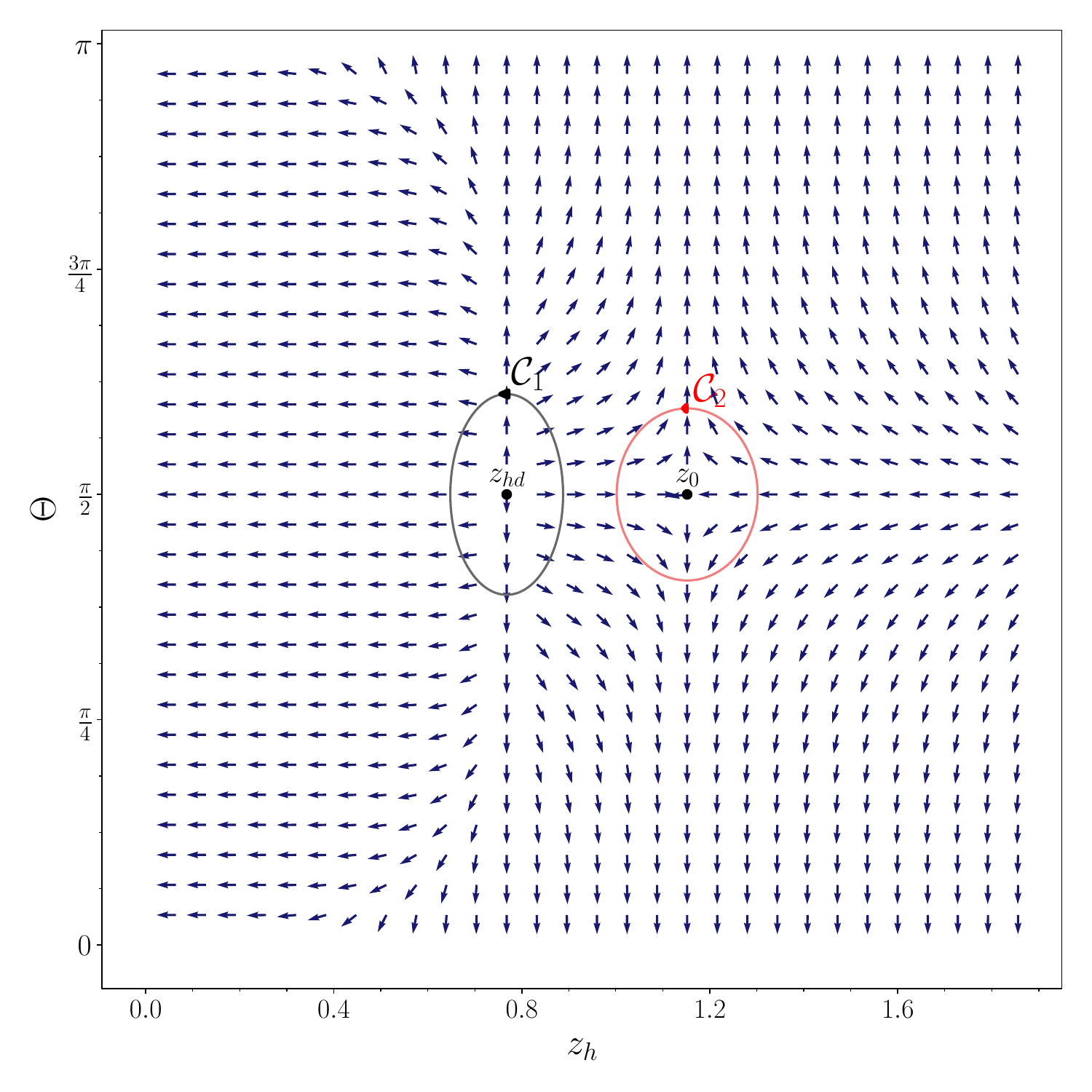}
            \caption{\(T > T_0\)}
            \label{fig: sw vector field above T0}
        \end{subfigure}\qquad
        \begin{subfigure}[t]{.45\linewidth}
            \includegraphics[width=\linewidth, height =.24\textheight]{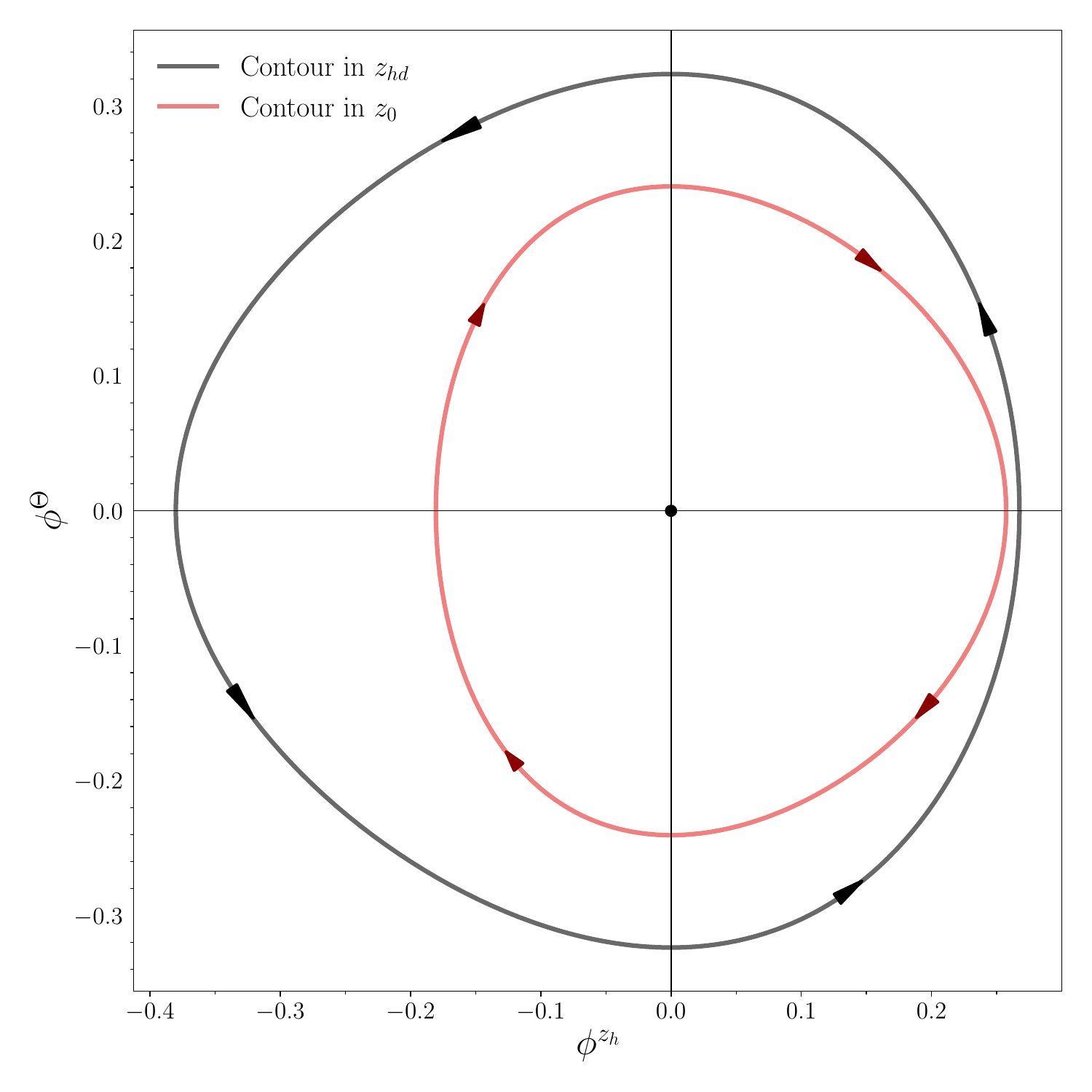}
            \caption{\(T > T_0\)}
            \label{fig: sw vector space above T0}
        \end{subfigure}
        \caption{
               (a), (c), (e) Normalized vector field obtained from \(\phi(z_h, \Theta)\) for the Schwarzschild-\(AdS_5\) within the Soft-wall model at increasing temperatures around \(T_0\). (b), (d), (f) The respective mapping of each contour in the vector space. The red contour encloses the \(z_0\) defect, while the gray contour encloses the \(z_{hd}\) defect.}
        
        \label{fig: sw vector field for different temperatures}
    \end{figure}
    }

    \subsection{Topological classification and Dynamics}

        As previously established, \(\phi^{z_h}\) possesses two roots, resulting in two topological defects located at (\(z_{hd}, \tfrac{\pi}{2}\)) and (\(z_0, \tfrac{\pi}{2}\)). Before analyzing the individual topological charges of these defects, let us determine the total charge \(W\) of the vector field by examining the behavior at the boundaries of the parameter space \(\p{z_h, \Theta}\).

        The \(\phi^\Theta\)-component exhibits the same behavior presented in Eq.\eqref{eq: phi component at edges}, pointing downward at the bottom (\(\Theta=0\)) and upward at top (\(\Theta=\pi\)). Thus, to establish the total topological charge, we must determine the \(z_h\)-components in the left (\(z_h\to0\)) and right (\(z_h\to\infty\)) edge.

        Expanding the field in the limit \(z_h\to0\) at the lowest non-trivial order, one finds 
        \begin{align}
            \phi^{z_h} &= \frac{L^3V}{\kappa^2}\p{-\frac{6}{z_h^5}+6\pi \frac{\bar T}{z_h^4}}+O(z_h^{-3})\nonumber\\
            &= \frac{L^3V}{2\kappa^2}\frac{3}{z_h^4}\p{-\frac{4}{z_h}+4\pi\bar T} + O(z_h^{-3})\,.
        \end{align}
        This is the same as was found in Eq.\eqref{eq: phi z_h schwarzschild}, indicating that in this limit of high temperatures, the energy scale of the SW is negligible compared to the thermal energy. So, the latter completely dominates the behavior of \(\phi\). Consequently, the field exhibits the same behavior, becoming static and pointing leftward along the entire left edge. Furthermore, the Hawking temperature root is directly obtained in this limit, evidencing the fact that in this limit, the black hole dominates near the boundary.

        Taking the expansion in the opposite limit, \(z_h\to\infty\), we find:  
        \begin{equation}
            \phi^{z_h} = \frac{L^3V}{\kappa^2}\p{-\frac{6\pi \bar T}{z_h^4}+e^{-c^2z_h^2}\p{\frac{8c^2\pi \bar T}{z_h^2}+O(z_h^{-3})}+O(z_h^{-5})}\,.
        \end{equation}
       
        As expected, \(\phi^{z_h}\) vanishes at infinity. However, due to the negative exponential in the second term, it is exponentially suppressed. Consequently, in contrast to what was presented in Sec.\ref{sec: Sec2}, the field approaches zero with negative values. So, the vector points inward at  $z_h\to\infty$. Thus, the contribution of this edge to the topological charge is \(-1/2\). Since the contributions from the remaining edges remain unchanged, the total charge is \(W=0\), which is conserved at any temperature \(T \ne 0\).  The change in the total topological charge serves as a signature showing that, in this regime, confinement dominates the behavior of the field in the large \(z_h\) limit.

        Let us analyze the behavior of the defects in the bulk. Note that \({\cal F}\) possesses two extrema points given by Eqs.\eqref{eq: hawking defect} and \eqref{eq: z0 minimum}, in addition to a static minimum at \(z_h\to\infty\). Since two local minima, considered to correspond to different values of \({\cal F}\) must be separated by a local maximum, the extrema located at the lowest value of coordinate $z$:  (\(z_0\) or \(z_{hd}\)) have to be a minimum. 
        
        In Fig.~\ref{fig: plot roots of phi}, we show the behavior of the finite minima of the free energy as a function of temperature. For low temperatures, the \(z_{hd}\) is large, hence, \(z_0\) is a minimum. At the temperature \(T_0\), given by
        \begin{equation}
            T_0 = \frac{1}{\pi z_0}\,,
        \end{equation}
        the two points coincide, and the free energy exhibits an inflection point. Above \(T_0\) the \(z_{hd}\) is a local minimum. It is important to note that this local minimum only becomes the global minimum at the Hawking-Page critical temperature \(T_c\).

        Thus, for temperatures below \(T_0\), Fig.~\ref{fig: sw vector field below T0}, the first defect corresponds to \(z_0\) and carries a positive topological charge (\(w=+1\)),   while the second at  \(z_{hd}\) possesses a negative charge (\(w=-1\)). As established in our previous analysis, these positive and negative charges correspond to counterclockwise and clockwise loops in the vector space, respectively,  Fig.~\ref{fig: sw vector space below T0}. This charge configuration indicates that,  in this temperature range, the black hole solution is unstable.

        For temperatures higher than \(T_0\), Fig.~\ref{fig: sw vector field above T0}, the defect at \(z_{hd}\) exhibits a positive topological charge (\(w=+1\)), while the defect at \(z_0\) acquires a negative charge (\(w=-1\)). As shown in Fig.~\ref{fig: sw vector space above T0}, the winding direction around each defect is reversed compared to the low temperature regime. This indicates that at these temperatures, the black hole solution exists as a stable minimum, while the SW configuration becomes unstable.

         Figure \ref{fig: sw vector field at T0} shows the vector configuration at \(T=T_0\).  At this temperature, the two defects coincide (\(z_0=z_{hd}\)), forming a degenerate point. This degeneracy indicates that \(\tfrac{\partial^2{\cal F}}{\partial z_h^2}=0\), and consequently, the topological charge for this merged defect vanishes (\(w=0\)). As can be seen in Fig.~\ref{fig: sw vector space at T0}, the corresponding contour in the vector space does not enclose the origin.

        Clearly, this point marks the exchange of topological charges and, consequently, the exchange of thermodynamic stability. Thus, it signals the temperature for the black hole formation, which corresponds to the creation of the plasma in the dual gauge theory. Let us examine this transition and the evolution of the topological charges in detail.

        By introducing the variables \(\uptau=T-T_0\) and \(x=z_h-z_0\) into the free energy \eqref{eq: sw off shell free energy}, and performing an expansion around the origin in terms of these variables (the details of this procedure, along with the explicit coefficients, are provided in Appendix \ref{Ap: 1}), one obtain the approximated free energy and the field component \(\phi^{z_h}\) as
        \begin{align}
            {\cal F}(x, \uptau)&\approx a_0 -\frac{1}{2} a_2 \uptau \,x^2 - \frac{1}{6}a_3 x^3\, \quad \mbox{with} \;\;a_0, a_2, a_3 >0\\ 
            \phi^{z_h}(x, \uptau)&=\frac{\partial {\cal F}}{\partial x}\approx x\p{-a_2\uptau- \frac{1}{2}a_3 x}\,.
        \end{align}

        The roots of the field are directly identified as \(x_1=0\) and \(x_2=-2\tfrac{a_2}{a_3}\uptau\). The sign of the topological charge is obtained by the second derivative of the free energy
        \begin{equation}
            \frac{\partial^2 {\cal F}}{\partial x^2} = -a_2\uptau -a_3 x
        \end{equation}
        evaluated at each stationary point. Consequently, the signs of the charges are given by
        \begin{equation}
            \eta_1 = \sign\p{-a_2\uptau}\quad \text{and}\quad \eta_2=\sign\p{a_2 \uptau}\,.
        \end{equation}

        Observe that the sum of the topological charges is always zero, ensuring charge conservation. As \(a_2\) is a positive constant, the local stability and topological charge sign are totally determined by \(\uptau\). For \(\uptau<0\), the first root, which corresponds to \(z_0\), acts as a stable minimum and carries the positive charge. As \(\uptau\to0\), the second derivative of the free energy vanishes, making the topological charge undetermined at \(\uptau =0 \), before it emerges with a negative charge for \(\uptau>0\). The exact opposite occurs for the second root, which corresponds to the \(z_{hd}\) defect. Furthermore, because \(x_2\) changes the sign at the critical point, this second root only represents a stable minimum when \(x_2<0\). This confirms that the black hole phase is thermodynamically stable only when its horizon satisfies \(z_{hd}<z_0\), locating it closer to the boundary.

\begin{figure}[ht]
            \begin{subfigure}[h]{0.45\textwidth}
                \includegraphics[width=1\linewidth]{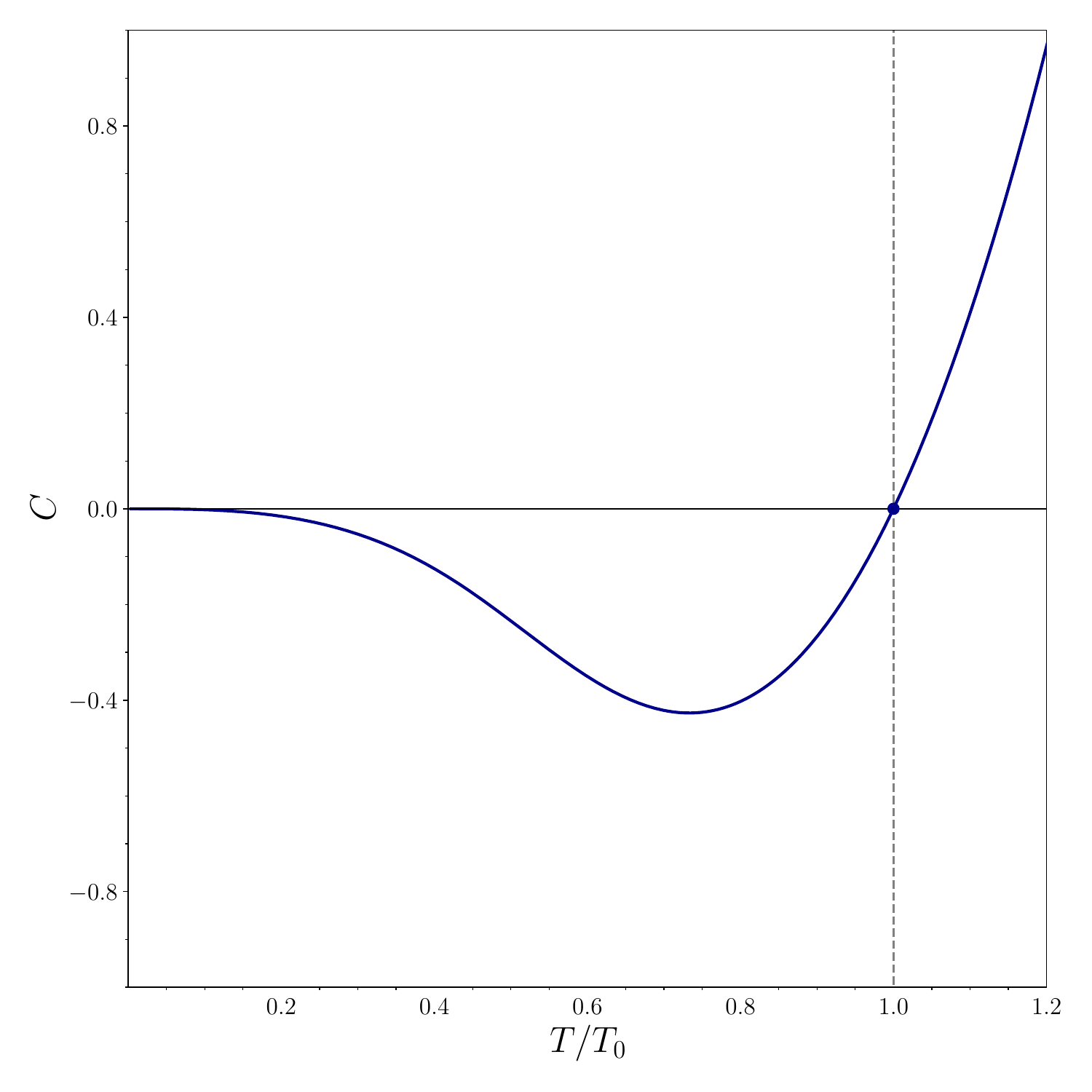}
                \caption{}
                \label{fig: sw heat capacity}
            \end{subfigure}
            \hfill
            \begin{subfigure}[h]{0.45\textwidth}
                \includegraphics[width=1.\linewidth]{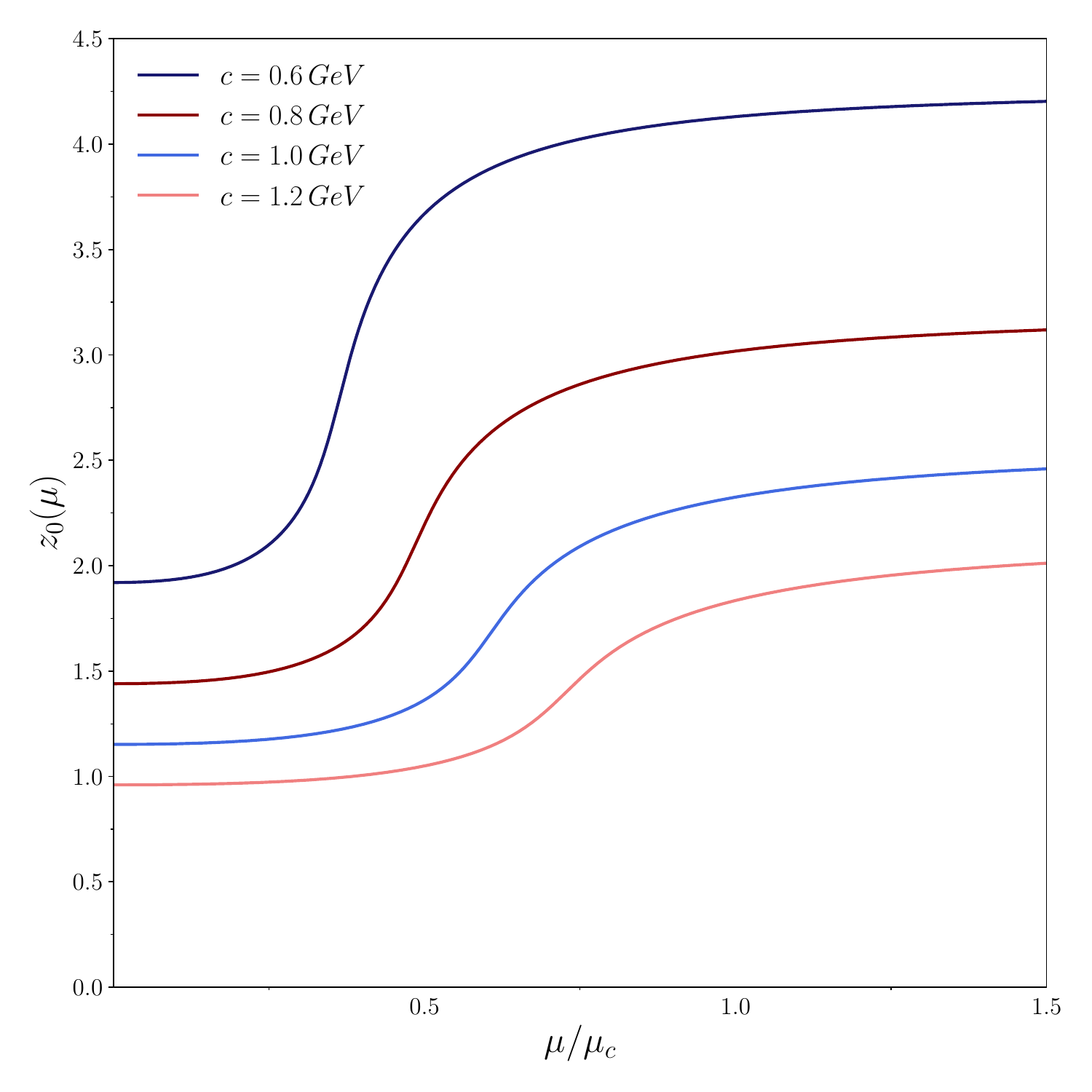}
                \caption{}
                \label{fig: z_0 as mu}
            \end{subfigure}
            \caption{
                 (a) The on-shell heat capacity for the soft-wall model as a function of temperature. The black dot marks the point where it becomes positive, indicating the black hole becomes locally stable. (b) Behavior of \(z_0\) as a function of the chemical potential \(\mu\) for various values of \(c\). The horizontal axis is normalized by the critical chemical potential at \(c=1.2\, GeV\).
                } 
                
        \end{figure}

        It is worth emphasizing that the sign of the heat capacity for the black hole, obtained as
        \begin{equation}
            C = -T\frac{\partial^2 {\cal F}}{\partial T^2 } = 4 T \frac{a_2^3}{a_3^2}\uptau\,,
        \end{equation}
      depends on the sign of  \(\uptau\). Consequently, \(C\) varies from negative (\(\uptau<0\)) to positive  (\(\uptau>0\)) values by passing continuously through zero as shown in Fig.~\ref{fig: sw heat capacity} for the on-shell heat capacity. This continuous behavior at vanishing heat capacity contrasts with that of standard black hole systems. 
    
        The black hole defect carries a charge of \(w=-1\) at low temperatures. Following the classification scheme of Ref.~\cite{Wei:2024gfz}, this places black hole into the \(W^{0-}\) topological class, in contrast to the \(W^{1+}\) class found in the pure AdS case. The shift in topological class is driven by the change in the background geometry, due to the confinement energy scale introduced by the dilaton.

        \subsection{The finite chemical potential case}

            At finite chemical potential, the off-shell free energy takes the form
            \begin{align}
                {\cal F} =& \frac{L^3 V}{\kappa^2}\frac{1}{z_h^4}e^{-c^2z_h^2}\left(\frac{1}{2} e^{c^2 z_h^2} \left(4 \pi  T z_h+3 \mu^2 z_h^2-3\right)-\frac{\mu ^2 \left(e^{c^2   z_h^2}-1\right) \left(8 \pi  T z_h+5 \mu ^2 z_h^2-6\right)}{c^2 \left(\mu ^2   z_h^2+2\right)}+\right.\nonumber\\&\left.c^2 z_h^2+c^4 z_h^4 e^{c^2 z_h^2} \text{Ei}\left(-c^2   z_h^2\right)+\frac{32-16 \pi  T z_h}{\mu ^2 z_h^2+2}+4 \pi  T z_h+2 \mu ^2 z_h^2-13 \right)\,.
            \end{align}
            Correspondingly, the \(z_h\)-component of the topological vector field becomes
            \begin{align}
                \phi^{z_h} =&\frac{L^3V}{\kappa^2}\frac{1}{z_h^5}e^{-c^2z_h^2}\p{-2+2\pi T + z_h^2\bar \mu^2} \left(4 c^2 {z_h}^2 \left(\frac{4}{\bar\mu ^2{z_h}^2+2}-1\right)+\right.\nonumber\\&\left.\frac{4 \bar\mu ^2 \left(e^{c^2 {z_h}^2}-1\right) \left(5 \bar\mu ^2 {z_h}^2+6\right)}{c^2 \left(\bar\mu ^2 {z_h}^2+2\right)^2}-3 e^{c^2 {z_h}^2}+\frac{24-14\bar\mu ^4 {z_h}^4}{\left(\bar\mu ^2 {z_h}^2+2\right)^2}\right)\,.\label{eq: zh phi to chemical potential case}
            \end{align}

            Setting the first factor in parentheses to zero directly yields the root \(z_{hd}\), which corresponds to the horizon position \eqref{eq: RN defect location}, as expected. As in the previous case, the root \(z_0\) arises from the last factor; however, in this case, \(z_0\) becomes strongly dependent on \(\bar \mu\) and can only be determined numerically. The profile of the \(z_0(\mu)\) as a function of \(\bar \mu\) is depicted in Fig.~\ref{fig: z_0 as mu}.

            Expanding \(\phi^{z_h}\) from Eq.\eqref{eq: zh phi to chemical potential case} in the limit \(z_h\to0\) one finds
            \begin{align}
                \phi^{z_h} \approx \frac{L^3V}{2\kappa^2}\frac{3}{z_h^4}\p{-\frac{4}{z_h}+4\pi\bar T+2\bar\mu^2 z_h +\frac{8}{3}c^2z_h}  + O(z_h^3)\,.
            \end{align}
            Once again, the thermal effects govern this regime, dictating that the topological vector points strictly leftward along the left boundary of the parameter space.

            Taking the expansion near the right boundary of the parameter space (i.e., \(z_h\to\infty\)) yields
            \begin{equation}
                \phi^{z_h}\approx\frac{L^3V}{\kappa^2}\p{-\frac{6\pi \bar T}{z_h^4}-\frac{3\bar \mu^2}{z_h^3}+e^{-c^2z_h^2}\p{-\frac{8c^2\pi \bar T}{z_h^2}+O\p{z_h^{-3}}}} +O\p{z_h^{-5}}\,.
            \end{equation}
            This demonstrates that the field approaches zero from the negative side, meaning the vector points inward in this limit.

            Since the topological vector field has the same asymptotic behavior as in the \(\bar \mu =0\) case, the total topological charge remains \(W=0\). This outcome is expected, since the field still contains two topological defects. This demonstrates that the total charge and the topological class are primarily determined by the dilaton or, in physical terms, by the confinement scale.

            It is worth noting that the local dynamics of the stability change is similar to those detailed in the previous section. The difference is that the phenomena occur at lower temperatures as the chemical potential increases. Since the black hole solution possesses a local topological charge of \(w=-1\) at low temperatures, it belongs to \(W^{0-}\) topological class, as anticipated in the previous paragraph.

\section{\label{sec: Sec4} Topological Hawking-Page transitions}

    While the topological vector field constructed from the off-shell free energy characterizes the local thermodynamic stability of the black hole branch, the effective temperature shifts the focus to the global phase structure. Explicitly studying the zero-free-energy condition (\({\cal F}=0\)), following the approach of \cite{Yerra:2022coh}, enables a topological classification of the global transitions between competing states.

    Considering the \(\mu=0\) case, we can obtain   an effective temperature for the Soft-wall model as
    \begin{equation}\label{eq: sw effective temperature}
        T_{eff} = \frac{\langle E\rangle}{S} = \frac{1}{\pi z_h}\p{4e^{-c^2z_h^2}-2}^{-1}\p{e^{-c^2z_h^2}\p{3+c^2z_h^2}-\frac{3}{2}+c^4z_h^4\text{Ei}\p{-c^2z_h^2}}\,.
    \end{equation}

    Redefining our topological vector field as
    \begin{equation}\label{eq: temperature vector field definition}
        \phi = \p{\frac{\partial T_{eff}}{\partial z_h}, - \frac{\cot{\Theta}}{\sin{\Theta}}}\,.
    \end{equation}
    the roots of \(\phi^{z_h}\) correspond to extrema of the effective temperature \eqref{eq: sw effective temperature}. Consequently, these topological defects indicate the critical points associated with the Hawking-Page phase transitions.

    Explicitly the \(z_h\)-component is given by
    \begin{equation}
        \phi^{z_h} = -\frac{\p{-6+3e^{c^2z_h^2}-4c^2z_h^2}\p{-2+2c^2z_h^2+e^{c^2z_h^2}\p{1+2c^4z_h^4}\text{Ei}\p{-c^2z_h^2}}}{4\p{-2+e^{c^2z_h^2}}^2z_h^2}\,.
    \end{equation}
    Setting the first factor in the numerator to zero  one gets    
    \begin{equation}
        z_0 = \frac{1}{c}\sqrt{\frac{1}{2}\p{-3-2k}}\,,
    \end{equation}
    which corresponds to the low-temperature transition identified in the previous section. The second factor does not possess an analytical root due to presence of the exponential integral function (Ei), but it represents the confinement/deconfinement phase transition point. We know from the numerical analysis that the critical temperature is \(T_c(0)=0.4917 c\); consequently, the associated defect is located at
    \begin{equation}
        z_{_{HP}}=\frac{1}{0.4917\pi c}\,.
    \end{equation}

\begin{figure}[ht]
            \begin{subfigure}[h]{0.47\textwidth}
                \includegraphics[width=1\linewidth]{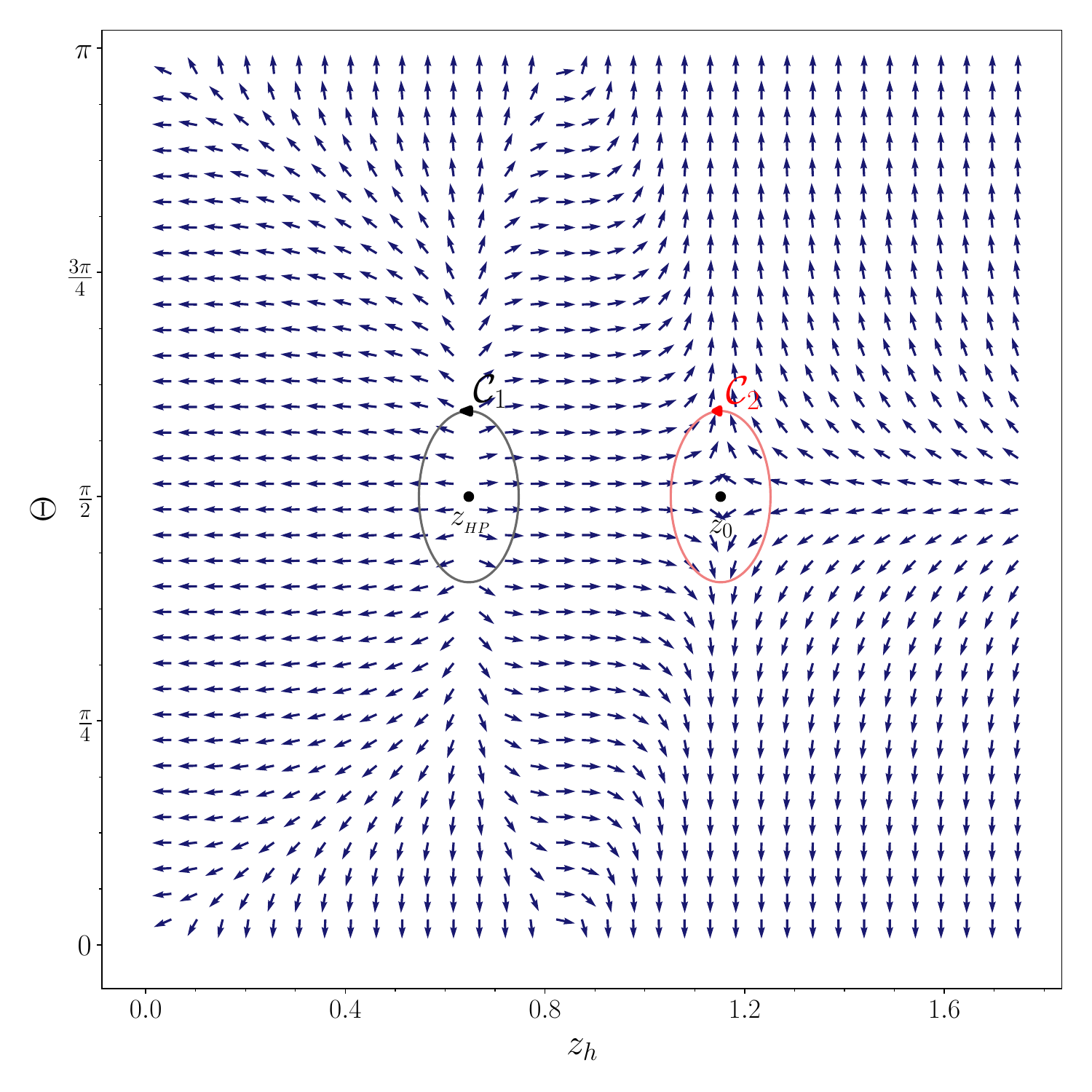}
                \caption{}
                \label{fig: sw temperature field}
            \end{subfigure}
            \hfill
            \begin{subfigure}[h]{0.47\textwidth}
                \includegraphics[width=1.\linewidth]{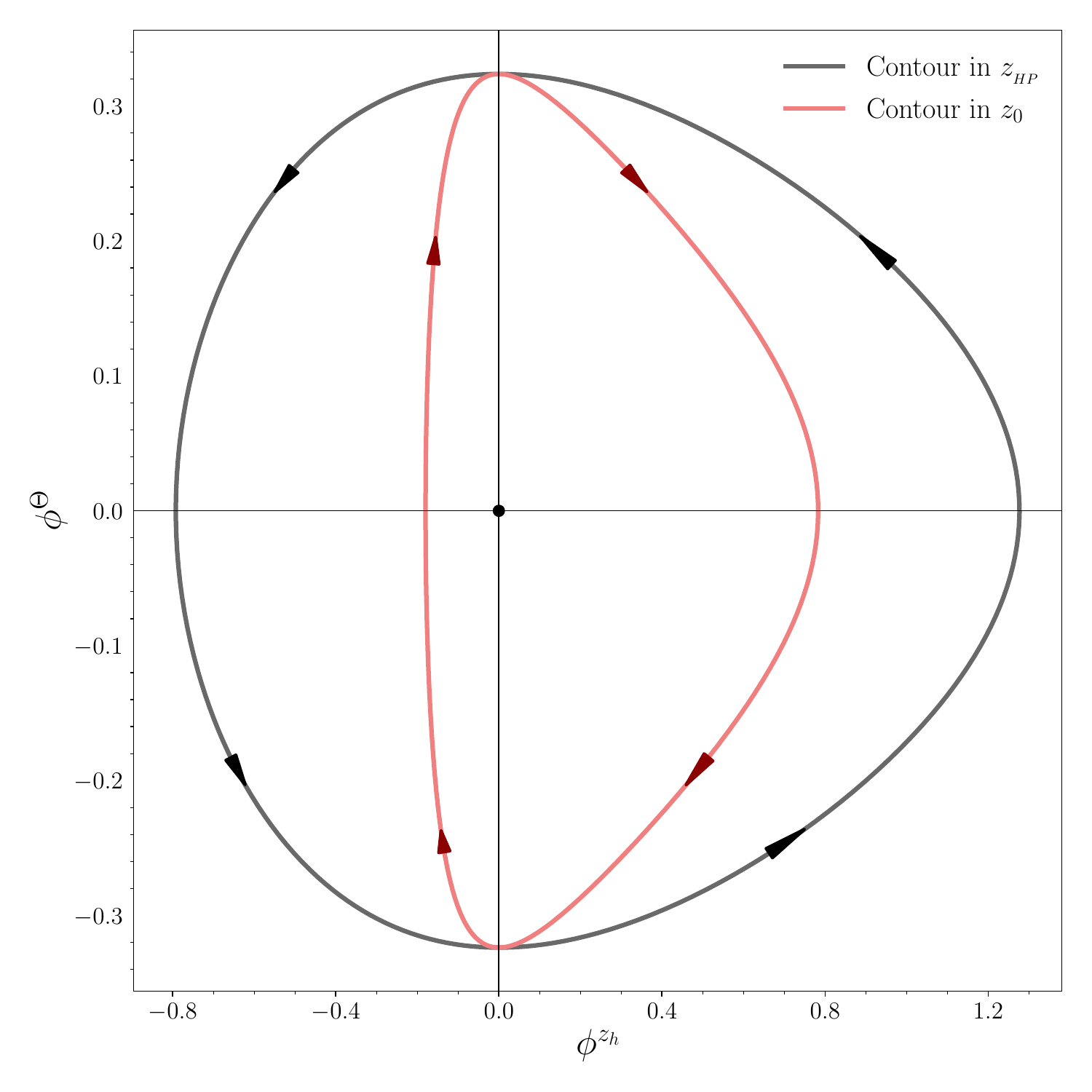}
                \caption{}
                \label{fig: sw temperature field vector space}
            \end{subfigure}
            \caption{
                (a) The normalized vector field obtained from effective temperature. The contour \({\cal C}_1\) encircles confinement/deconfinement topological defect (\(z_{_{HP}}\)), whereas \({\cal C}_2\) encloses the defect located at \(z_0\). (b) The mapping of the vector field's winding direction along the respective contours.
               }
                
        \end{figure}

    Figure \ref{fig: sw temperature field} shows the vector field in parameter space, featuring the two defects located at \(\p{z_{_{HP}}, \tfrac{\pi}{2}}\) and \(\p{z_{0}, \tfrac{\pi}{2}}\). Similar to the previous analysis, the sign of the topological charge for each defect is determined by the sign of the second derivative of \(T_{eff}\) evaluated at the respective defect. Thus, taking this second derivative at \(z_{_{HP}}\) yields 
    \begin{equation}
        \left.\frac{d^2T_{eff}}{dz_h^2}\right|_{z_h=z_{_{HP}}}= K_1(c) \quad\text{with}\quad K_1>0
    \end{equation}
    indicating that the charge for this defect is \(w_1=+1\). This result is confirmed by the counterclockwise winding of the vector field along the contour enclosing it in Fig.\ref{fig: sw temperature field vector space}.

        Considering the second defect at \(z_0\) we find
        \begin{equation}
            \left.\frac{d^2T_{eff}}{dz_h^2}\right|_{z_h=z_{0}}= - K_2(c) \quad\text{with}\quad K_2>0 \,,
        \end{equation}
        leading to a topological charge of \(w_2=-1\). This result is confirmed by the clockwise loop along the contour in the Fig.\ref{fig: sw temperature field vector space}. Clearly, these charges indicate that the defects are located at a minimum and a maximum of the effective temperature. Let us now extract the physical significance of this topological information.

        For simplicity, let us introduce the shifted variable \(x=z_h-z_i\) , where \(z_i\in \{z_{_{HP}}, z_0\}\) denotes the location of a  topological defect. Starting with the Hawking-Page transition defect at  \(z_{_{HP}}\)(i.e., \(x=z_h-z_{_{HP}}\)), performing an expansion of \(T_{eff}\) around \(x=0\) leads to
        \begin{align}
            T_{eff}\p{x} &\approx T_{eff}\p{x=0}+ \left.\frac{\partial T_{eff}(x)}{\partial x}\right|_{x=0} \,x +\frac{1}{2}\left.\frac{\partial^2 T_{eff}(x)}{\partial x^2}\right|_{x=0}x^2+O\p{x^3}\nonumber\\
            &\approx T_c(0) + \frac{1}{2} K_1 x^2 + O\p{x^3}\,,\label{eq: effective temperature expansion around hp}
        \end{align}
        where the linear term vanishes because \(x=0\) is a minimum of $T_{eff}$.  

        Expanding the entropy \eqref{eq: sw entropy} around the same point, we obtain
        \begin{align}
            S(x)&\approx S(x=0) +\left.\frac{d S}{d x}\right|_{x=0} x + O\p{x^2}\nonumber\\
            &\approx S_0 -S_1 x
        \end{align}
        where \(S_0, S_1>0\). Substituting this relation into Eq.\eqref{eq: effective temperature expansion around hp}, we find
        \begin{equation}
            T_{eff}(S)\approx T_c\p{0} +\frac{K_1}{2S_1^2}(S-S_0)^2\,.
        \end{equation}

        This equation defines an upward-facing parabola in the \(T\)-\(S\) plane, with its absolute minimum located precisely in \(T_c\p{0}\). Physically, this demonstrates that no black hole state within this stable manifold can achieve a zero-free-energy equilibrium with the thermal vacuum at a temperature lower than \(T_c\p{0}\). This finding rigorously confirms that \(T_c\p{0}\) is a stable global minimum of the effective temperature \(T_{eff}\), aligning with the standard expectations for a genuine Hawking-Page phase transition \cite{Yerra:2022coh}.

        Turning to the transition defect at \(z_0\) (where \(x=z_h-z_0\)), the effective temperature expansion takes the form
        \begin{equation}\label{eq: effective temperature expansion around z0}
            T_{eff}(x) \approx T^* - \frac{1}{2} K_2 x^2 + O(x^3)\,,
        \end{equation}
        where \(T_{eff}\p{x=0}=T^*\). It is straightforward to verify from \eqref{eq: sw entropy} that \(z_0\) is a minimum of the entropy. Thus, expanding the entropy to second order around this point yields
        \begin{equation}
            S = - s_0 + \frac{1}{2}s_2 x^2 + O\p{x^3}.
        \end{equation}
        Substituting this expression back into Eq.\eqref{eq: effective temperature expansion around z0}, we obtain the effective temperature as a linear function of entropy
        \begin{equation}
            T_{eff}(S) \approx T^* - \frac{ K_2}{s_2}(S+s_0)\,.
        \end{equation}

        The emergence of a linear expansion with a negative slope is a direct consequence of the vanishing first derivative of the entropy at this critical point. This indicates that \(T_{eff}(S)\) lacks a minimum and decreases monotonically with the entropy, characterizing a non-physical transition. This conclusion is in perfect agreement with our earlier thermodynamic analysis, where we established that the free-energy minimum at \(z_0\) possesses negative entropy and constitutes a pathological state.
    
        This analysis demonstrates that while a positive topological charge (\(w=+1\)) corresponds to a genuine thermodynamic transition, a negative charge \(w=-1\) unambiguously identifies a thermodynamically unstable and nonphysical crossing.

\section{\label{sec: Conclusão} Conclusion}

    In this work, we have applied Duan's \(\phi\)-mapping formalism to a phenomenological holographic model, with the intention of studying the confined/deconfined phase transition under the topological point of view.

    Following the approach proposed in \cite{Wei:2022dzw}, where the black hole solutions emerge as defects of a vector field derived from off-shell thermodynamic free-energy, we first classified the \(AdS_D\) black hole solutions with planar boundary (\(R^{D-2}\times S^1\)). We showed that these geometries belong to the \(W^{1+}\) topological class. This classification arises because they possess only one stable state at any given temperature, which appears as a single positive defect in the topological field for both the Schwarzschild and Reissner-Nordström cases.

    This scenario implies that the plasma in the dual theory is always the dominant state, which is inconsistent with QCD predictions; therefore, it is necessary to employ a model that introduces a confinement energy scale to break the conformal symmetry. The model analyzed in this work is the soft-wall model, which introduces an energy scale by inserting a dilaton field into the gravitational action (neglecting the backreaction), modifying the infrared limit of the dual gauge theory.

    Here, our main result emerges: this modification effectively redefines the thermal AdS state and, consequently, alters the boundary conditions of the thermodynamic parameter space $\{z_h, \Theta\}$ at   (\(z_h\to\infty\)). Because the boundary is fundamentally changed, the system is forced into a new topological class \(W^{0-}\), shifting from a trivial conformal background to a dynamically rich confining one. Thus, we have shown that the topological charge acts as a robust, macroscopic invariant that mathematically captures the emergence of the confinement scale in bottom-up holographic models.

    Remarkably, an analysis of the local stability dynamics reveals a degenerate point at the temperature \(T_0\). Here, the topological charges vanish, and the defects swap signs, effectively establishing the black hole, and by extension, the plasma formation as a locally stable minimum. It is important to emphasize that for any non-zero temperature, the total topological charge remains \(W=0\), as it is a conserved quantity. A detailed discussion of these dynamics is presented in Sec.~\ref{sec: Sec3}.

    To study the global transitions, we construct a new the topological vector field using an effective temperature that satisfies the condition \({\cal F}=0\). In this scenario, the confinement/deconfinement transition is characterized by a positive topological charge. This result is expected for a first-order transition and agrees with the study of  Hawking-Page transitions in Ref.~\cite{Yerra:2022coh}. Furthermore, we demonstrated that this positive charge is associated with a genuine physical transition, whereas the negative charge marks a virtual, nonphysical transition, as discussed in Sec.~\ref{sec: Sec4}.

    Therefore, we have shown that the introduction of the confinement scale places this deformed AdS geometry into the \(W^{0-}\) topological class, even at a finite chemical potential. It is known that other geometries used to holographically describe the confinement/deconfinement transition, such as the Schwarzschild AdS with compact boundary, also belong to this class. Thus,  further studies investigating other holographic frameworks, such as Einstein-Dilaton models like \cite{Gursoy:2007cb,Gursoy:2007er,Ballon-Bayona:2021tzw}      or full top-down approaches like \cite{Sakai:2004cn}, would be interesting in order to understand whether this \(W^{0-}\) class serves as a topological signature of the confinement.

\begin{acknowledgments}
    The authors are partially supported by CNPq --- Conselho Nacional de Desenvolvimento Científico e Tecnológico, by FAPERJ --- Fundação Carlos Chagas Filho de Amparo à Pesquisa do Estado do Rio de Janeiro, and by  Coordenação de Aperfeiçoamento de Pessoal de Nível Superior --- Brasil (CAPES), Finance Code 001.
\end{acknowledgments}

\appendix

    \section{\label{Ap: 1} The off-shell free energy expansion}

        Considering the off-shell free energy written in terms of the variables \(x=z_h-z_0\) and \(\uptau = T-T_0\), we perform a Taylor expansion near \(x=0\)
        \begin{align}
            {\cal F}(x, \uptau)&\approx {\cal F}\p{x=0, \uptau} +\left.\frac{\partial }{\partial x}{\cal F}(x, \uptau)\right|_{x=0} \,x +\frac{1}{2}\left.\frac{\partial^2 }{\partial x^2}{\cal F}(x, \uptau)\right|_{x=0} \,x^2 +\frac{1}{6}\left.\frac{\partial^3 }{\partial x^3}{\cal F}(x, \uptau)\right|_{x=0} \,x^3  \nonumber\\
            &\approx l_0(\uptau) +\frac{1}{2}l_2\p{\uptau}x^2 + \frac{1}{6}l_3\p{\uptau}x^3 +O\p{x^4}\label{eq: ap-free energy taylor expasion}
        \end{align}
        where the first derivative vanishes because \(x=0\) is an extremum of the free energy. Now, let us expand in \(\uptau\) each coefficient around \(\uptau=0\) in lowest order, the zeroth-order coefficient becomes
        \begin{equation}\label{eq: ap-l0}
            l_0(\uptau)\approx l_0(\uptau=0)\approx c^4\p{\text{Ei}\p{\frac{3}{2}+k}+\frac{5}{k\p{6+4k}}}=a_0\,,
        \end{equation}
        where \(k={\cal W}_{-1}(-\tfrac{3}{4}e^{-3/2})\).

        For the second-order coefficient, the constant term \(l_2\p{0}\) vanishes because \(\p{x=0, \uptau=0}\) is an inflection point. Thus, expanding \(l_2(\uptau)\) to linear order
        \begin{align}
            l_2(\uptau) &\approx l_2(\uptau=0) + \left.\frac{d}{d\uptau}l_2(\uptau)\right|_{\uptau=0}\uptau \nonumber\\
            &\approx-24c^5\pi\frac{\sqrt{-6-4k}\p{1+k}}{k\p{3+2k}^2 }\uptau=-a_2\uptau\,.\label{eq: ap-l2}
        \end{align}

        Finally, for the third-order coefficient, it suffices to retain only the zeroth-order term in \(\uptau\), yielding
        \begin{equation}\label{eq: ap-l3}
            l_3(\uptau)\approx l_3\p{0}\approx-96c^7\frac{\sqrt{2\p{-3-2k}}\p{1+k}}{k\p{3+2k}^2}=-a_3\,.
        \end{equation}

        Therefore, substituting Eqs. \eqref{eq: ap-l0}, \eqref{eq: ap-l2} and \eqref{eq: ap-l3} into the free energy expansion \eqref{eq: ap-free energy taylor expasion}, we obtain
        \begin{equation}
            {\cal F}\p{x,\uptau} \approx a_0 - \frac{1}{2}a_2\uptau \,x^2- \frac{1}{6}a_3 x^3 + O(x^4)\,.
        \end{equation}





\bibliography{mybib} 

\end{document}